\documentclass[
pdftex,
twocolumn,
epjc3]{svjour3}          % twocolumn

\RequirePackage[T1]{fontenc}

\smartqed  % flush right qed marks, e.g. at end of proof

\RequirePackage{graphicx}
\RequirePackage{mathptmx}      % use Times fonts if available on your TeX system
\RequirePackage{flushend}
\RequirePackage[numbers,sort&compress]{natbib}
\RequirePackage[colorlinks,citecolor=blue,urlcolor=blue,linkcolor=blue]{hyperref}

\usepackage{amssymb}
\usepackage{amsmath}
\usepackage{appendix}
\usepackage{graphicx}% Include figure files
\usepackage{dcolumn}% Align table columns on decimal point
\usepackage{bm}% bold math
\usepackage{gensymb}
\usepackage{graphicx}
\usepackage[caption=false, justification=justified]{subfig}
\usepackage{microtype}
\usepackage{boxhandler}
\usepackage{hyphenat}
\usepackage{xcolor}

\def\nonubb{$0\nu\beta\beta$}

\def\ppc{PPC}  
\def\MJ{{\sc Majorana}}             %Majorana project name
\def\DEM{{\sc Demonstrator}}             %Demonstrator in small caps

\def\ITEP{National Research Center ``Kurchatov Institute'' Institute for Theoretical and Experimental Physics, Moscow, 117218 Russia}
\def\JINR{Joint Institute for Nuclear Research, Dubna, 141980 Russia} 
\def\lbnl{Nuclear Science Division, Lawrence Berkeley National Laboratory, Berkeley, CA 94720, USA}
\def\lanl{Los Alamos National Laboratory, Los Alamos, NM 87545, USA}
\def\queens{Department of Physics, Engineering Physics and Astronomy, Queen's University, Kingston, ON K7L 3N6, Canada}
\def\uw{Center for Experimental Nuclear Physics and Astrophysics, and Department of Physics, University of Washington, Seattle, WA 98195, USA}
\def\unc{Department of Physics and Astronomy, University of North Carolina, Chapel Hill, NC 27514, USA}
\def\duke{Department of Physics, Duke University, Durham, NC 27708, USA}
\def\ncsu{Department of Physics, North Carolina State University, Raleigh, NC 27695, USA}	
\def\ornl{Oak Ridge National Laboratory, Oak Ridge, TN 37830, USA}

\def\pnnl{Pacific Northwest National Laboratory, Richland, WA 99354, USA}
\def\sdsmt{South Dakota School of Mines and Technology, Rapid City, SD 57701, USA}
\def\usc{Department of Physics and Astronomy, University of South Carolina, Columbia, SC 29208, USA}
\def\usd{Department of Physics, University of South Dakota, Vermillion, SD 57069, USA}  
\def\ut{Department of Physics and Astronomy, University of Tennessee, Knoxville, TN 37916, USA}
\def\tunl{Triangle Universities Nuclear Laboratory, Durham, NC 27708, USA}
\def\mpi{Max-Planck-Institut f\"ur Physik, M\"unchen, 80805, Germany}
\def\tum{Physik-Department, Technische Universit\"at, M\"unchen, 85748 Germany}
\def\williams{Physics Department, Williams College, Williamstown, MA 01267, USA}
\def\iu{Department of Physics, Indiana University, Bloomington, IN 47405, USA}
\def\iuceem{IU Center for Exploration of Energy and Matter, Bloomington, IN 47408, USA}

\journalname{Eur. Phys. J. C}
\begin{document}

\title{\texorpdfstring{$\alpha$}{Alpha}-event Characterization and Rejection in Point-Contact HPGe Detectors}

\author{
I.J.~Arnquist \thanksref{pnnl} 
\and
F.T.~Avignone~III \thanksref{usc}\thanksref{ornl}
\and
A.S.~Barabash \thanksref{ITEP}
\and
C.J.~Barton \thanksref{usd}
\and
F.E.~Bertrand \thanksref{ornl}
\and 
E. Blalock \thanksref{ncsu}\thanksref{tunl}
\and 
B.~Bos\thanksref{unc}\thanksref{tunl} 
\and 
M.~Busch\thanksref{duke}\thanksref{tunl}
\and 
M.~Buuck\thanksref{SLAC}\thanksref{uw}
\and 
T.S.~Caldwell\thanksref{unc}\thanksref{tunl}	
\and 
Y-D.~Chan\thanksref{lbnl}
\and 
C.D.~Christofferson\thanksref{sdsmt} 
\and 
P.-H.~Chu\thanksref{lanl} 
\and 
M.L.~Clark\thanksref{unc}\thanksref{tunl} 
\and 
C.~Cuesta\thanksref{CIEMAT}\thanksref{uw}
\and 
J.A.~Detwiler\thanksref{uw}	
\and 
A.~Drobizhev\thanksref{lbnl} 
\and 
T.R.~Edwards\thanksref{lanl}\thanksref{usd}
\and 
D.W.~Edwins\thanksref{usc} 
\and 
F.~Edzards\thanksref{mpi}\thanksref{tum}
\and 
Y.~Efremenko\thanksref{ut}\thanksref{ornl}
\and 
S.R.~Elliott\thanksref{lanl}
\and 
T.~Gilliss\thanksref{APL}\thanksref{unc}\thanksref{tunl}
\and 
G.K.~Giovanetti\thanksref{williams}  
\and 
M.P.~Green\thanksref{ncsu}\thanksref{tunl}\thanksref{ornl}
\and 
J.~Gruszko\thanksref{email}\thanksref{unc}\thanksref{tunl}
\and I.S.~Guinn\thanksref{unc}\thanksref{tunl} 
\and V.E.~Guiseppe\thanksref{ornl}	
\and C.R.~Haufe\thanksref{unc}\thanksref{tunl}	
\and R.J.~Hegedus\thanksref{unc}\thanksref{tunl} 
\and R.~Henning\thanksref{unc}\thanksref{tunl}
\and D.~Hervas~Aguilar\thanksref{unc}\thanksref{tunl} 
\and E.W.~Hoppe\thanksref{pnnl}
\and A.~Hostiuc\thanksref{uw} 	
\and I.~Kim\thanksref{lanl} 
\and R.T.~Kouzes\thanksref{pnnl}
\and A.M.~Lopez\thanksref{ut}	
\and J.M. L\'opez-Casta\~no\thanksref{ornl} 
\and E.L.~Martin\thanksref{duke}\thanksref{tunl}	
\and R.D.~Martin\thanksref{queens}	
\and R.~Massarczyk\thanksref{lanl}		
\and S.J.~Meijer\thanksref{lanl}
\and S.~Mertens\thanksref{mpi}\thanksref{tum}		
\and J.~Myslik\thanksref{lbnl}		
\and T.K.~Oli\thanksref{usd}  
\and G.~Othman\thanksref{Hamburg}\thanksref{unc}\thanksref{tunl} 
\and W.~Pettus\thanksref{iu}\thanksref{iuceem}	
\and A.W.P.~Poon\thanksref{lbnl}
\and D.C.~Radford\thanksref{ornl}
\and J.~Rager\thanksref{ARA}\thanksref{unc}\thanksref{tunl}	
\and 
A.L.~Reine\thanksref{unc}\thanksref{tunl}	
\and K.~Rielage\thanksref{lanl}
\and N.W.~Ruof\thanksref{uw}	
\and B.~Sayk\i\thanksref{lanl} 
\and S.~Sch\"{o}nert\thanksref{tum}
\and M.J.~Stortini\thanksref{lanl} 
\and D.~Tedeschi\thanksref{usc}		
\and R.L.~Varner\thanksref{ornl}  
\and S.~Vasilyev\thanksref{JINR}	
\and J.F.~Wilkerson\thanksref{unc}\thanksref{tunl}\thanksref{ornl}
\and M.~Willers\thanksref{tum}
\and C.~Wiseman\thanksref{uw}		
\and W.~Xu\thanksref{usd} 
\and C.-H.~Yu\thanksref{ornl}
\and B.X.~Zhu\thanksref{JPL}\thanksref{lanl} 
}

\thankstext[$\star$]{email}{Corresponding author, e-mail: jgruszko@unc.edu}
\thankstext{SLAC}{Present address: SLAC National Accelerator Laboratory, Menlo Park, CA 94025, USA}
\thankstext{CIEMAT}{Present address: Centro de Investigaciones Energ\'{e}ticas, Medioambientales y Tecnol\'{o}gicas, CIEMAT 28040, Madrid, Spain}
\thankstext{APL}{Present address: Applied Physics Laboratory, Johns Hopkins University, Laurel, MD 20723, USA}
\thankstext{Hamburg}{Present address: Universit\"at Hamburg, Hamburg, 20146, Germany}
\thankstext{ARA}{Present address: Applied Research Associated, Raleigh, NC 27615, USA}
\thankstext{JPL}{Present address: Jet Propulsion Laboratory, California Institute of Technology, Pasadena, CA 91109, USA}

%\collaboration{{\sc{Majorana}} Collaboration}
%\noinstitute

\institute{
\pnnl \label{pnnl}
\and
\usc \label{usc}
\and
\ornl \label{ornl}
\and
\ITEP \label{ITEP}
\and
\usd \label{usd}
\and
\ncsu \label{ncsu}
\and
\tunl \label{tunl}
\and
\unc\label{unc}
\and
\duke \label{duke}
\and
\uw \label{uw}
\and
\lbnl \label{lbnl}
\and
\sdsmt\label{sdsmt}
\and
\lanl \label{lanl}
\and
\mpi \label{mpi}
\and
\tum \label{tum}
\and
\ut \label{ut}
\and
\williams \label{williams}
\and
\queens \label{queens}
\and 
\iu \label{iu}
\and 
\iuceem\label{iuceem}
\and
\JINR \label{JINR}
}

\date{Received: date / Accepted: date}

\maketitle

\begin{abstract}
P-type point contact (\ppc) HPGe detectors are a leading technology for rare event searches due to their excellent energy resolution, low thresholds, and multi-site event rejection capabilities. We have characterized a \ppc\ detector's response to $\alpha$ particles incident on the sensitive passivated and p$^+$ surfaces, a previously poorly-understood source of background. The detector studied is identical to those in the \MJ\ \DEM\ experiment, a search for neutrinoless double-beta decay (\nonubb) in $^{76}$Ge. $\alpha$ decays on most of the passivated surface exhibit significant energy loss due to charge trapping, with waveforms exhibiting  a delayed charge recovery (DCR) signature caused by the slow collection of a fraction of the trapped charge. The DCR is found to be complementary to existing methods of $\alpha$ identification, reliably identifying $\alpha$ background events on the passivated surface of the detector. We demonstrate effective rejection of all surface $\alpha$ events (to within statistical uncertainty) with a loss of only 0.2\% of bulk events by combining the DCR discriminator with previously-used methods. The DCR discriminator has been used to reduce the background rate in the \nonubb\ region of interest window by an order of magnitude in the \MJ\ \DEM\, and will be used in the upcoming LEGEND-200 experiment.
\end{abstract}

\section{\label{sec:intro}Introduction}
\subsection{\texorpdfstring{$\alpha$}{Alpha} Backgrounds in Neutrinoless Double Beta Decay Searches}
The discovery of neutrinoless double-beta decay (\nonubb) would indicate that neutrinos are Majorana particles and that lepton number is not conserved. It could also provide information on the absolute mass scale of the neutrino, the source of the neutrino's mass, and the origin of the matter/anti-matter asymmetry of the universe \cite{Dolinski_2019, Barabash_2019, PhysRevD.25.2951}. This radioactive decay would occur only rarely; sensitivity limits from current experiments indicate that the half-life is over $10^{26}$\,yrs \cite{KLZ2016, GERDA_final}. Detecting it therefore requires large-mass experiments with extremely low background rates and the best possible energy resolution. 

P-type point contact (\ppc) High-Purity Ger\-ma\-ni\-um (HP\-Ge) detectors \cite{luk89} are a key technology for rare event searches, capable of detection thresholds below 1\,keV and full-width at half-maximum (FWHM) energy resolutions of 0.12\% (2.5\,keV) at the \nonubb\ region-of-interest of 2039\,keV \cite{MJD2019}. They have been used to search for low-energy nuclear recoils from external sources in dark matter searches, \cite{Abgrall2017} and have been proposed for use in coherent neutrino-nuclear scattering experiments \cite{Barbeau2007}. They are also a leading technique for \nonubb\ searches, where the detector itself is the source \cite{MJD2019, GERDA2019}. The \nonubb\ experiments using \ppc\ detectors have the lowest background rates of any of the currently-operating experiments \cite{GERDA2019}. 

A troubling source of backgrounds for \nonubb\ experiments is the decay of radon isotopes and their progeny, particularly $^{222}$Rn, on or near the surface of the detectors. Long-lived $\alpha$-emitters such as $^{210}$Po are the most concerning; the 138 day half-life of this isotope makes the use of timing correlation-based rejection prohibitively inefficient. If its decay is supported by the decay of $^{210}$Pb, the associated 22 year half-life will make the $^{210}$Po rate approximately constant throughout the life of the experiment. Since the $\alpha$ particle is emitted with 5.304\,MeV and that energy is easily degraded when passing through even a thin layer of inactive material, it can appear in the region of interest (ROI) around the $^{76}$Ge \nonubb\ Q-value of 2.039\,MeV. 

Because the range of a 5.304\,MeV $\alpha$ particle in germanium is less than 20\,$\mu$m \cite{SRIM}, all $\alpha$ decays from sources outside the detector are considered surface events in HPGe detectors. There is no evidence for bulk $\alpha$ events, which would have to originate from contaminants in the HPGe material. If these events did occur, they would not contribute to the \nonubb\ ROI, since their full energy would be deposited inside the detector.

In \ppc\ detectors, the majority of the surface is covered in a 1-2\,mm thick lithium-diffused dead layer forming the n$^+$ contact \cite{Aguayo2013}, and is therefore completely insensitive to $\alpha$ decays. The remaining sensitive surfaces are those of the central p$^+$ contact region, formed by boron implantation in the germanium bulk, and the passivated surface, a $\sim0.1\,\mu$m layer of amorphous germanium or silicon. The dimensions of these regions depend on the chosen detector geometry; two common designs are the Mirion \footnote{formerly Canberra} BEGe \ppc\ and ORTEC \ppc\ detectors. The ORTEC \ppc\ detectors used in the \MJ\ \DEM\ have a passivated surface that covers an entire circular face of the detector, with area of approximately 30\,cm$^2$ per detector. 

The response of a \ppc\ detector to events near the passivated surface is difficult to predict, and depends on the detector manufacturing process. Charge trapping has been observed on similar surfaces in other HPGe detector geometries \cite{Abt2017}, but the charge collection properties can differ depending on the surface treatment and field configuration. In the \MJ\ \DEM\ detectors, events have been observed in which $\alpha$ particles originating on this surface are significantly degraded in energy, appearing in the \nonubb\ ROI, as discussed in Ref.~\cite{MJD2019}. We performed a dedicated study of $\alpha$ interactions on this surface, leading to more reliable models of the $\alpha$ energy spectrum and the distinctive pulse-shape characteristics of these signals. 

\subsection{\label{sec:DCR_intro} The Delayed Charge Recovery Effect}

\begin{figure}
  \centering
 \includegraphics[trim={0.6in 3in 1in 3.2in},clip, width=\linewidth]{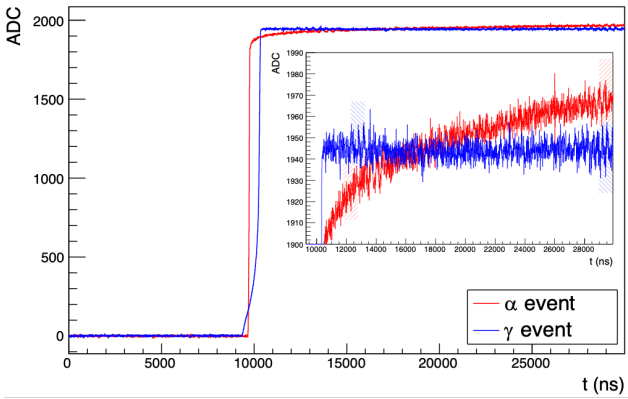}
  \caption{Baseline-removed and pole-zero corrected waveforms from PONaMa-1 events, taken with the TUBE scanning system. Both the bulk $\gamma$ event \textit{(in blue)} and the surface $\alpha$ event, taken with the source incident at r$=-7.5$\,mm \textit{(in red)}, have 2615\,keV of energy. The waveform regions used to calculate the DCR parameter for each waveform, determined as described in Sec.~\ref{ssec:PSD_DCR}, are indicated by the shaded boxes, shown in the same color as the waveform they correspond to.}
 \label{fig:alphaWF}
\end{figure}

Based on the characteristics of $\alpha$ interactions, it appears that charge mobility is drastically reduced on or near the passivated surface. Therefore, a fraction of the charge from these interactions is slowly released on the timescale of waveform digitization, leading to a measurable increase in the slope of the tail of the waveform. 

An offline digital filter can be used to identify these events, allowing for the efficient rejection of passivated surface $\alpha$ events \cite{MJD2019}. The goal of such a filter is to detect the presence of slow charge collection occurring after the bulk charge collection has been completed. This filter is used to calculate the Delayed Charge Recovery (DCR) parameter. In a waveform that has been corrected for the electronic response function, this appears as a positive slope of the tail, as seen in Fig.~\ref{fig:alphaWF}.

The delayed component of the surface $\alpha$ signal can be modeled by considering the motion of electrons and electron-holes near the passivated surface. Computational models of charge drift in germanium detectors have shown that on the surface of the detector, the carrier drift velocities may be 10 to 100 times lower than in the detector bulk \cite{Mullowney2012}. In the ``slow surface drift" model, some fraction of the charge carriers are driven to the passivated surface, with the remainder of the carriers being collected normally through the bulk. This behavior could be caused by the presence of a net charge on the passivated surface, or by the self-repulsion of charge carriers in the dense charge cloud created by an $\alpha$ interaction; these cases can be distinguished, to some extent, by the differing detector response to such events as a function of $\alpha$ interaction position.

Alternatively or in addition to the slow surface drift behavior, some fraction of the charge carriers may be trapped in a few-$\mu$m-thick region just below the passivated surface and then slowly re-released, dominating the observed slow charge collection behavior. If the detector passivation process causes any damage to the germanium crystal structure (for example, via sputtering causing crystal dislocations), this would create a narrow higher-trapping region near the surface. This would not affect bulk charge collection, but would cause incomplete charge collection in events occurring on or near the surface. In this case, which we term the ``near-surface trapping" model, the remainder of the charge carriers can be collected promptly, as they are in bulk events, collected slowly, as in the slow surface drift model, or trapped at the detector surface due to passivated surface charge build-up, contributing negligibly to the signal.

In all of these cases, part of the energy of the event appears as a normal, fast pulse, and the remainder of the charge is collected slowly. Depending on which charge carriers are affected, these models produce different predictions of event energy and DCR as the position of the $\alpha$ interaction on the passivated surface changes. A dedicated scan along a radial path of a \ppc\ detector's passivated surface with a collimated $\alpha$-emitting source can not distinguish between the two causes of delayed charge, but can reveal which charge carriers are being affected. The near-surface trapping and slow surface drift effects may both be present, with the radial behavior governed by the dominant effect. 

The expected \nonubb\ efficiency of the DCR pulse-shape discriminator can be determined from single-site $^{228}$Th events, as discussed in Ref.~\cite{MJD2019}. Its leakage (i.e.~the fraction of $\alpha$ events that are misidentified as signal-like events), however, requires \textit{a priori} knowledge of the number of $\alpha$ events in the detector. Since the DCR varies as a function of interaction position, the leakage can also vary as a function of position, requiring additional knowledge of the spatial $\alpha$ event distribution. A dedicated surface scan using a spectroscopic-grade $\alpha$ source is one of the few reliable ways to measure the DCR discriminator's leakage.   

\begin{figure}
  \centering
 \includegraphics[trim={1in 0 1in 0},clip, width=\linewidth]{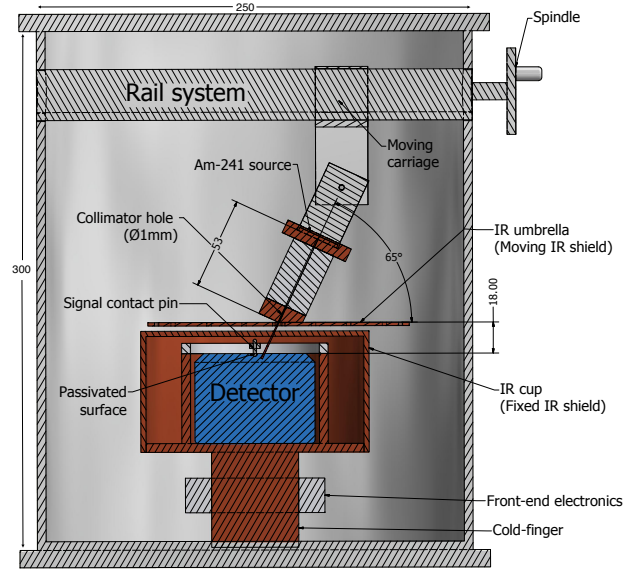}
 \caption{A simplified cross-sectional view of the TUBE scanner, showing key dimensions in millimeters. Details of the detector cup, front-end electronics, and cold-finger are removed for clarity. Both the detector holder (unlabeled) and the IR cup are held at the ground potential, with an insulating spacer placed between the detector holder and the n$^+$ surface of the detector, which is at high voltage.}
 \label{fig:TUBE}
\end{figure}

\section{\label{sec:exp}Experimental Setup and Calibration}
\subsection{\label{ssec:tube}The TUBE Scanner}
The TUM (Technische Universit\"at, M\"unchen) Upside-down BEGe (TUBE) scanner is a custom-built cryostat first made to study the backgrounds in GERDA due to surface interactions on the p$^+$ electrode and insulating surface of Mirion BEGe \ppc\ detectors \cite{AgostiniThesis}. It allows a \ppc\ detector's passivated surface to be scanned with a collimated source. Key aspects of the scanner design, following minor modifications needed for use with an ORTEC PPC detector, can be seen in Fig.~\ref{fig:TUBE}. 

Measurements of the detector's $\alpha$ response were taken with an open 40\,kBq $^{241}$Am $\alpha$ spectrometry source with a vendor-determined $\alpha$ energy peak full-width at half-maximum (FWHM) of 20\,keV. The source was mounted in a 1\,mm-diameter collimator. The entire 68.9\,mm diameter of the detector's passivated surface was scanned, save for a 6-mm ``blind spot" on the detector surface that is occluded by signal electronics components. The p$^+$ contact is only partially occluded by the signal electronics. The 0$\degree$ and 180$\degree$ positions along the scanned diameter are distinguished by assigning them positive and negative radial positions. The source beam had a 1.8\,mm-diameter spot size on the detector surface. 

When calculated from the background-subtracted data as described in Sec.~\ref{ssec:eff}, the average $\alpha$ event rate (including only source positions where the beam is not occluded) was $17.9\pm0.5$\,mHz. This is consistent with the 17.6\,mHz expected rate derived from the collimator geometry and manufacturer-cited source strength. A muon veto system was used to reduce backgrounds from cosmogenic muons at high energy. Events from both the detector and the muon veto system were recorded with a Struck SIS3302 digitizer, sampling at 100\,MHz with a trace length of $30\,\mu$s. Offline analysis was used to reject multi-site events, reducing the rate of background $\gamma$ events, as described in Sec.~\ref{ssec:PSD}.

\begin{figure}
  \centering
 \includegraphics[width=\linewidth]{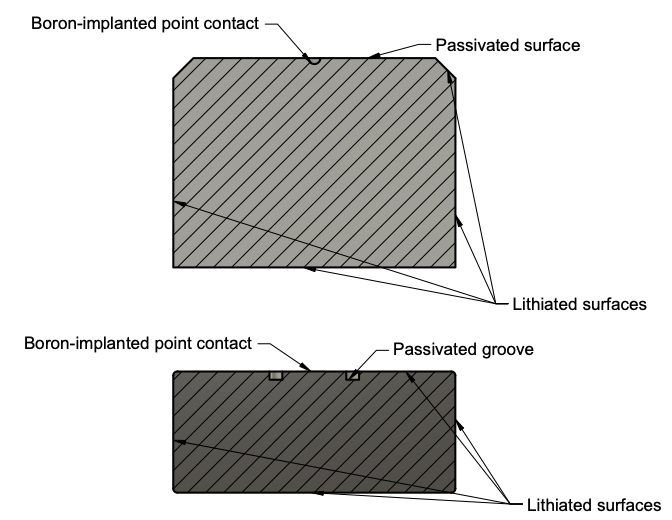}
 \caption{Diagrams of the commonly-used PPC detector geometries from ORTEC \textit{(top)} and Mirion \textit{(bottom)}.}
 \label{fig:detectors}
\end{figure}

The scanned detector, named PONaMa-1 (\textbf{P}PC from \textbf{O}RTEC made from \textbf{Na}tural \textbf{Ma}terial), was produced by ORTEC. Its fabrication was identical and geometry similar to the $^{76}$Ge-enriched  detectors used in the \MJ\ \DEM. The hemispherical p$^+$ contact was made by boron implantation and is $0.3\,\mu$m thick. The passivated surface covers nearly an entire circular face of the crystal, and has a radius of 30\,mm. The dimensions and other key characteristics of the detector are given in Table~\ref{tab:PONaMA_specs}, and a diagram of the detector is shown in Fig.~\ref{fig:detectors}.

See \ref{app:scanner} for more details about the source, collimator, and scanner.  

\begin{table}
\centering
\begin{tabular}{p{5cm} | l}
\hline
\multicolumn{2}{c}{PONaMa-1 Properties} \\
\hline
Diameter & 68.9\,mm \\ 
Height & 52.0\,mm \\ 
n$^+$ Dead Layer Thickness & 1.2\,mm \\
Passivated Surface Diameter & 60\,mm \\
p$^+$ Contact Diameter & 3.2\,mm \\
p$^+$ Contact Depth & 2.0\,mm \\
Capacitance & 1.8\,pF \\
Depletion Voltage & 850\,V \\
Leakage Current & 10\,pA \\
Resolution at 1332\,keV & 2.05\,keV \\
\end{tabular}
\caption{Dimensions and operating parameters of the PONaMa-1 \ppc\ detector. The dimensions were determined by the detector manufacturer. The n$^+$ dead layer thickness, capacitance, depletion voltage, leakage current, and resolution were determined by the detector manufacturer and then confirmed with independent measurements conducted as part of the \MJ\ \DEM\ detector characterization campaign \cite{Mertens_2015}.}
 \label{tab:PONaMA_specs}
\end{table}

\subsection{\label{ssec:scan}Scanning and Calibration Measurements}
The detector was operated with a bias voltage of 1050\,V, which is 200\,V above the depletion voltage of the crystal. Measurements were taken every 1.5\,mm along a radius with intermediate 0.75\,mm points near the p$^+$ contact and passivated surface edge. Each measurement lasted 24 hours. Several multi-day runs were taken to study the stability of the system, and a subset of scanning positions were repeated non-contiguously to study the long-term stability of the DCR parameters.

The data analyzed herein correspond to two deployments of the detector. The first deployment was a set of scans taken over 217 days of continuous operation. During this time, the detector was kept biased, cooled, and under vacuum. After these 217 days, the detector was warmed and the cryostat was opened. The source position was adjusted to give a higher $\alpha$ event rate, and the detector was put back into operation. The data from this second deployment are used only for studies of the detector response stability, and are not included in other analyses.

The high ambient background rate allowed the energy to be calibrated independently for each data set, without the need for dedicated calibration runs. Additional runs with $^{232}$Th and $^{228}$Th sources were also used to confirm the energy estimation performance and calibrate pulse shape discrimination parameters. For a description of the calibration procedure and discussion of the detector energy scale stability, see \ref{app:calibration}.

\subsection{\label{ssec:PSD}Rising-Edge-Based Pulse Shape Discrimination Parameters} 

\begin{figure*}[htb]
  \centering
 \includegraphics[trim={2in 0in 2in 0},clip, angle=90, origin = c, width=\linewidth]{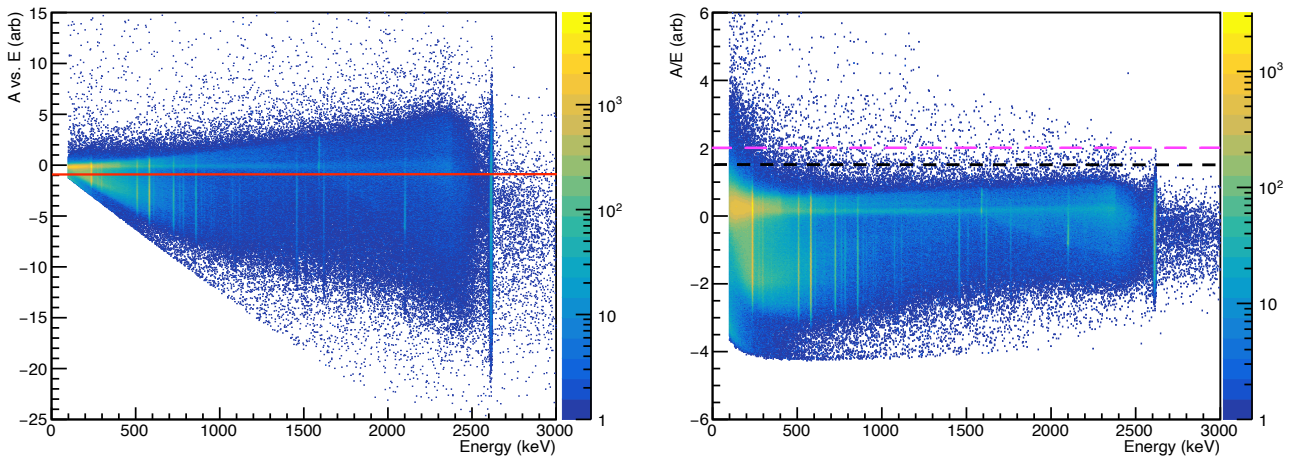}
 \caption{The distribution of A vs.~E \textit{(left)} and A/E \textit{(right)} values with respect to energy is shown for calibration events after the muon veto is applied. The color scale indicates the number of events. A vs.~E is used to identify multi-site $\gamma$ background events; all events with A vs.~E less than -1 (indicated by the red line) are rejected. A/E is used to identify near-p$^+$ contact events; 99.9\% of bulk events with energies between 1 and 2.63\,MeV have A/E less than 2 (indicated by the dashed magenta line), and 99.8\% have A/E less than 1.5 (indicated by the dashed black line).}
 \label{fig:AE}
\end{figure*}

In \ppc\ detectors, the charge drift time varies dramatically as a function of the interaction position. This prompt charge drift forms the shape of the rising edge of the waveform. Pulse shape discriminators that are sensitive to this portion of the signal can provide information including the interaction location and whether interactions occurred at multiple sites within the crystal. The rising edge is also sensitive to changes in the electric field within the bulk of the crystal. In the present analysis, one algorithm, A vs.~E, is used for multi-site $\gamma$ background event rejection; another, A/E, is used to identify near-point-contact $\alpha$ events. These algorithms measure the peak value of the current pulse ($A$), normalized with respect to the event energy ($E$). They differ in how the normalization is applied, but are otherwise similar. Their implementation is described in Refs.\,\cite{Agostini2013, MJD_avse}. A third algorithm, which directly measures the drift time of the pulse, is used to track the stability of the detector's electric field over time. 

$\alpha$ events are intrinsically single-site, since the particles have short range in germanium. Ambient background events, which are primarily high-energy $\gamma$ Compton interactions, are often multi-site. These backgrounds can be reduced via the use of a multi-site event cut. The most common techniques rely on the peak amplitude of the derivative of the waveform ($A$, which corresponds to the maximum signal current) as a function of energy ($E$), which is reduced in multi-site events relative to its value in single-site events. Applying a lower bound to this rising-edge-based discrimination parameter preferentially removes multi-site events. In this work, we use the A vs.~E algorithm described in Ref.\,\cite{MJD_avse}. 

A similar parameter, A/E \cite{Agostini2013}, can be used to reject (or select) near-point-contact events. The fast drift times of these events increase their peak current values for a given energy. Therefore, an upper-bound cut on A/E preferentially cuts events that occur near the $p^+$ contact. See Refs.~\cite{GERDA2017} and \cite{AgostiniThesis} for more discussion of this approach. 

Fig~\ref{fig:AE} shows the A vs.~E and A/E distributions with respect to energy in a $^{228}$Th calibration data set. A vs.~E, which has higher signal efficiency at low energy, is used to reject multi-site $\gamma$ events, as in the analysis employed in the \MJ\ \DEM\ \cite{MJD_avse}. A/E is used to identify near-point-contact events, as in the GERDA experiment, where it is used to reject $\alpha$ and $\beta$ surface events \cite{GERDA2017}. A/E can also be used to identify multi-site $\gamma$ events, but it is not used for that purpose in this analysis.

A vs.~E and A/E were calibrated based on two $^{228}$Th calibrations, one taken at the start of data-taking and the second taken following an observed gain shift. Both are calibrated such that the lower-bound threshold cut on the parameter accepts 90\% of events in the $^{208}$Tl double-escape peak. See Refs.~\cite{MJD_avse} and~\cite{GERDA2017} for more details.  

 The upper-bound bulk acceptance values of A/E are calibrated using runs with no $\gamma$ or $\alpha$ calibration sources present. They are based on the acceptance of events with energies between 1 and 2.63\,MeV after the application of a muon veto cut and a basic pile-up cut. No multi-site event rejection cut is applied. The A/E distribution is normalized so that the 99\% acceptance value occurs at a value of 1. With this normalization, the 99.9\% acceptance value of A/E is found to be $2.00\pm0.05$. A cut of A/E\,$>1.5$, which is used to select near-point-contact events in the $\alpha$ energy analysis (see Sec.~\ref{ssec:energy}), is found to accept 99.8\% of events. As seen in Fig.~\ref{fig:AE}, employing a similar near-point-contact event cut in A vs.~E would require an energy-dependent upper threshold. Therefore, A vs.~E is not used to select near-point-contact events in this analysis. The energy-dependent width of the A vs.~E distribution is due to charge cloud diffusion; this effect is corrected, to first order, by the energy normalization used for the A/E parameter. Additional discussion of this effect can be found in Ref.~\cite{MJD_AvsE_Talk}, and will be included in an upcoming publication from the \textsc{Majorana} Collaboration.

To study the bulk event pulse shape stability in the detector, as described in \ref{app:stability}, we also compute the ``$t_0$ to $t_{50}$" drift time of each pulse. The start time of the pulse, $t_0$, is found by applying a trapezoidal filter with an integration time of 1.5\,$\mu$s and a collection time of 1\,$\mu$s, identifying the time at which the filtered pulse passes a fixed level threshold of 5\,ADC counts, and then correcting for the trapezoidal filter timing offset. The relatively high threshold was chosen to give stable results under changing noise conditions. The 50\% rise time of the pulse, $t_{50}$, is found by linearly interpolating between sampled points and identifying the first time at which the waveform crosses 50\% of its maximum value. The difference between the start time and 50\% rise of the pulse is taken to be the drift time. The $t_0$ parameters were chosen to provide sensitivity to the sharp initial rise of near-point-contact events without being subject to noise, and the $t_{50}$ point is used to avoid the degradation of the drift time parameter by the DCR effect itself.

\subsection{\label{ssec:PSD_DCR}Tail-Based Pulse Shape Discrimination Parameters} 

\begin{figure}
  \centering
 \includegraphics[trim={0.2in 0 0.5in 0},clip, width=\linewidth]{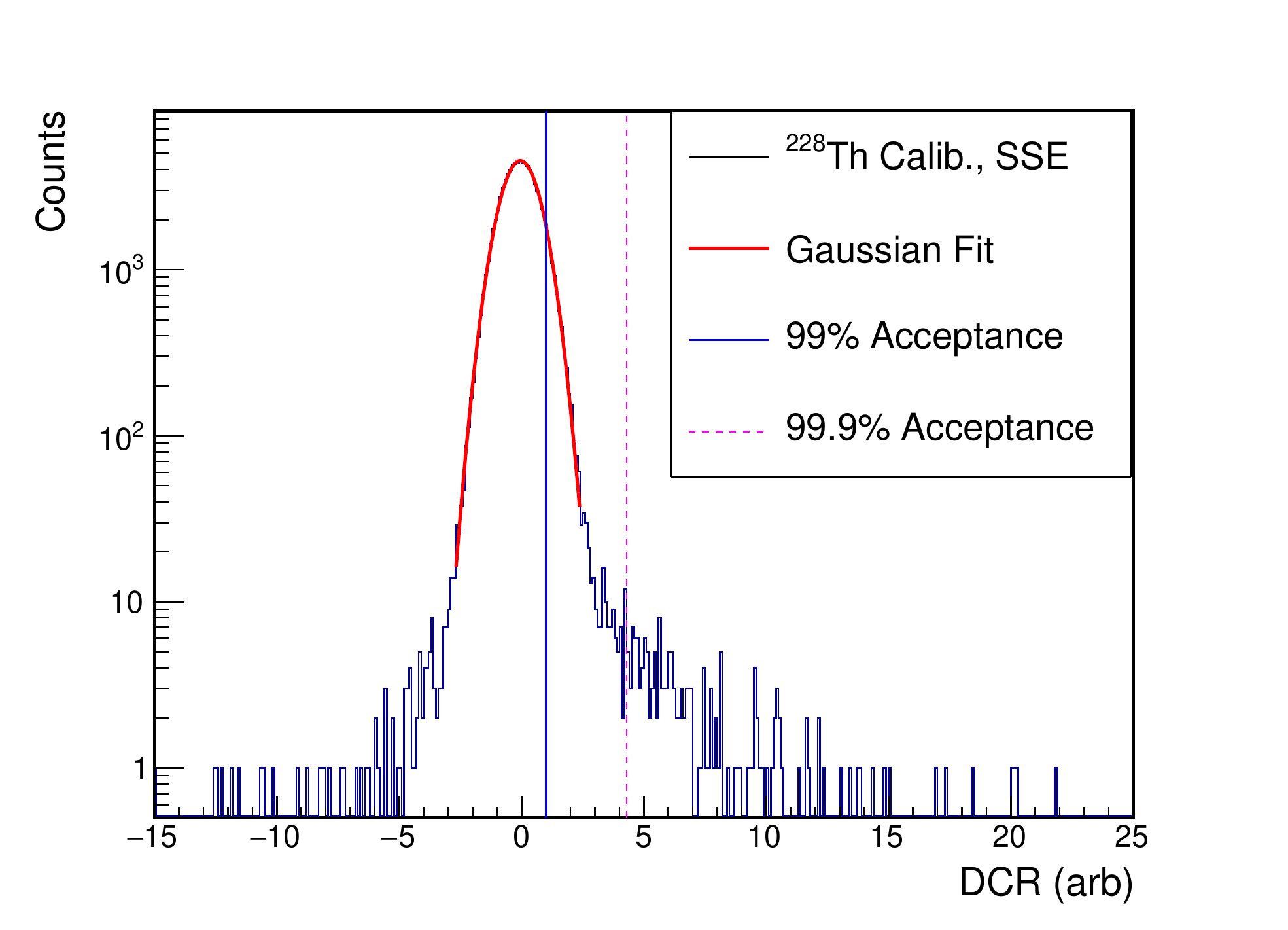}
 \caption{The DCR distribution in a $^{228}$Th calibration data set. Non-muon single-site events with energies between 1 and 2.63\,MeV are used to calibrate the DCR parameter. The centroid of a Gaussian fit (shown in red) is used to shift the tail slope $\delta$, and the distribution is normalized to the 99\% acceptance value (in blue). The 99.9\% acceptance value (shown as a dashed violet line) is also provided for reference.}
 \label{fig:DCR_calib}
\end{figure}

In a \ppc\ detector operated above its depletion voltage, bulk events exhibit a sharp, step-like transition from the rising edge of the waveform to its tail, as seen in the $\gamma$ event shown in Fig.~\ref{fig:alphaWF}. All of the charge collection occurs promptly and ends at that transition. Passivated surface $\alpha$ events, on the other hand, exhibit additional slow charge collection after the prompt signal has ended. This DCR effect can be measured via pulse shape discrimination parameters that are sensitive to the shape of the waveform tail.  

To calculate the DCR parameter, the resting baseline, taken to be the average ADC value of the first $5\,\mu$s (500 samples) of each waveform, is subtracted from the waveform. Then the shaping effect of the pre-amplifier, which adds a $44.390\,\mu$s pole-zero decay constant to the signal, is deconvolved from each waveform. Second-order shaping effects remain, but their effects are minimized by selecting waveform regions for the DCR calculation that are slightly delayed relative to the end of the waveform rise. 

The tail slope ($\delta$) of each waveform is found from a two-point slope calculation. This approach minimizes the parameter computation time per waveform. It can be expressed as:
\begin{equation}
    \delta = \frac{\overline{V(t_2)}-\overline{V(t_1)}}{(t_2-t_1)} 
    \label{eqn:delta}
\end{equation} 
where $\overline{V(t_i)}$ is the average value of the waveform (in ADC) in the $1\,\mu$s time window starting at $t_i$. The first point, $t_1$, is 2\,$\mu$s after the 97\% rise point of the waveform, and $t_2$ is set so that the second average is over the final 1\,$\mu$s of the waveform. In this case, 30\,$\mu$s traces are recorded, so $t_2$ is set to be 29\,$\mu$s after the start of the waveform. The delayed charge recovery continues beyond the end of the digitization window. See Fig.~\ref{fig:alphaWF} (right inset) for an example. 

This unnormalized tail slope $\delta$ gives the rate of delayed charge collection. $E_{d}$, the energy that is collected as delayed charge during the digitization window, is given by:
\begin{equation}
    E_d = at_t\delta
    \label{eqn:delayedE}
\end{equation}
where $a$ is the linear energy calibration constant and $t_t$ is the length of the tail of the waveform ($18\,\mu$s, in this case). Since the digitization window is limited to 30\,$\mu$s, this measured delayed charge does not include all of the delayed charge from each event. Given the limitations of the digitizer used and the high ambient background rate, the true total charge recovered cannot be measured in TUBE.

The calculation of the normalized DCR value is based on calibration data sets, which are contiguous sets of $^{228}$Th or $^{232}$Th calibration runs. Following the application of a muon veto and basic pile-up cut, the distribution of values of $\delta$ for single-site calibration events with energies between 1 and 2.38\,MeV is fit with a Gaussian peak. The fit range is set to exclude the high- and low-DCR tails. The 99\% and 99.9\% acceptance values of DCR are calculated from this same event population, including the events in the tails of the distribution. 

The normalized DCR value is then given by:
\begin{equation}
    \text{DCR} = \frac{\mu - \delta}{\sigma_{99}}
    \label{eqn:dcr}
\end{equation} 
where $\mu$ is the centroid of the fit, $\delta$ is the tail slope, calculated as in Eqn~\ref{eqn:delta}, and $\sigma_{99}$ is the 99\% acceptance value of the shifted $\delta$ distribution. 

After parameter calibration, the mean value of DCR is 0 and 99\% of bulk events have DCR\,$< 1$. The 99.9\% acceptance value is $4.3\pm0.9$. The DCR parameter was recalibrated during each $^{232}$Th or $^{228}$Th calibration run, with calibration occurring every 2 to 4 weeks. The uncertainty on the 99.9\% acceptance cut value captures the fluctuations in the high DCR tail of the distribution over the course of these calibration runs. The normalized DCR distribution for calibration run events is shown in Fig.~\ref{fig:DCR_calib}. 

We note that the delayed charge component of the waveform tail observed is not truly linear; it is more accurately fit by an exponential rise. Fitting an exponential function to each waveform, however, adds computational complexity to the waveform processing without measurably improving $\alpha$ event rejection. Therefore the faster first-order DCR algorithm given by Eqns.~\ref{eqn:delta} and \ref{eqn:dcr} is studied here. The full exponential fit can be used to find the time constant of delayed charge collection, as discussed in Sec.~\ref{ssec:dcr}.

\section{\texorpdfstring{$\alpha$}{Alpha} Event Response}\label{sec:alpha_meas} 
\subsection{\texorpdfstring{$\alpha$}{Alpha} Event Energy}\label{ssec:energy}
By studying the measured energy of fixed-energy $\alpha$ events at a variety of positions on the detector surface, we can probe the mechanism of energy degradation as discussed in Section~\ref{sec:siggen}. These measurements can also be used to create more accurate spectral models of $\alpha$-emitting background sources in low-background experiments such as the \MJ\ \DEM. 

$^{241}$Am $\alpha$ events distributed in a broad peak are observed for every source position incident on the passivated surface or p$^+$ contact. When the source beam is partially or entirely incident on the p$^+$ contact, an $\alpha$ peak in the spectrum appears at nearly the full energy of the emitted $\alpha$. For positions incident on the passivated surface, the $\alpha$ events are degraded in energy, with the peak energy and width varying with radius. 

Across most of the detector surface, muon veto, pile up rejection, and multi-site rejection cuts are applied, and the remaining $\alpha$ peak in the energy spectrum is fit with a Gaussian function. For positions with radii smaller than 6\,mm, an additional cut requiring A/E $> 1.5$ is used to identify near-point-contact events, and the $\alpha$ peak fitting function contains an additional component accounting for the peak's low energy tail. The same approach is used for all small-radius ($|r|<6$\,mm) data sets. The low-energy tail contains a majority of the events in data sets with $|r|\leq4.5$\,mm, so for these data sets, an additional estimated energy range of the observed $\alpha$ events is given. This energy range is given in lieu of the mean position of the Gaussian for the smallest-radius data sets ($|r|<3$\,mm), where the Gaussian+tail model fit fails. See \ref{app:energy} for details of the cuts applied in each case and the peak fitting procedure. 

\begin{figure}
  \centering
 \includegraphics[width=\linewidth]{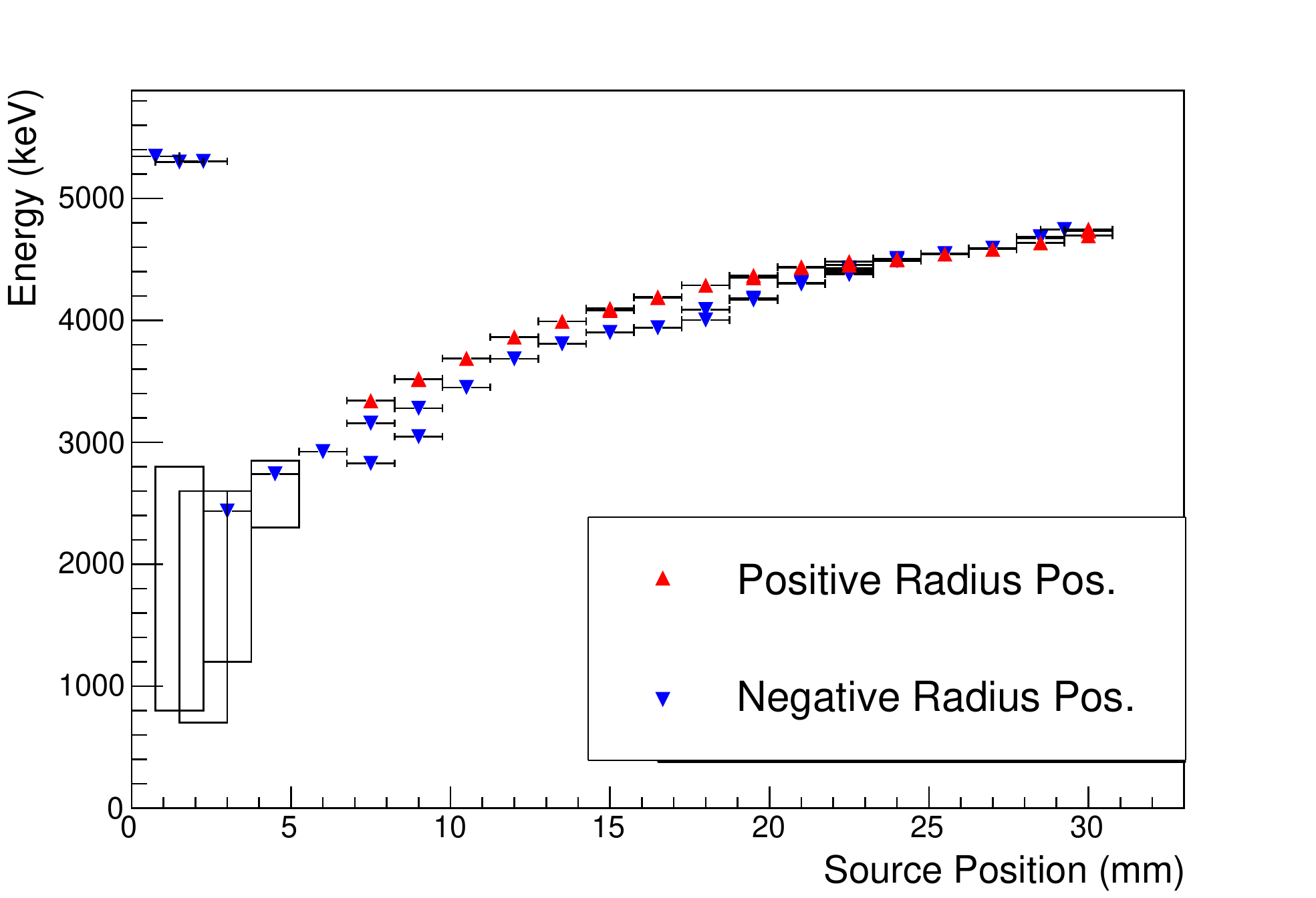}
 \caption{The centroids of the $\alpha$ energy peaks in each data set; certain positions were studied in multiple data sets. Negative (positive) radius positions are indicated by blue downward-pointing (red upward-pointing) triangles. For scanning positions with significant low-energy tailing, the black box depicts the estimated full energy range of $\alpha$ events. At positions that are partially or completely incident on the point-contact, an additional peak appears at nearly the full incident $\alpha$ energy. The vertical error bars (not visible) depict the uncertainty in the peak position from the maximum-likelihood fit of the peaks, and the horizontal errors depict the 0.75\,mm estimated uncertainty of the source position. The observed instability of the $\alpha$ peak energies is discussed in \ref{app:stability}.}
 \label{fig:EvR}
\end{figure}

All of the peak energies of the fits to the $\alpha$ energy spectra are depicted in Fig.~\ref{fig:EvR}. For positions with the source incident on the passivated surface, the full width at half maximum (FWHM) of the $\alpha$ peaks ranges from 50\,keV, at large-radius positions, to 240\,keV, at the smallest-radius positions. At the p$^+$ contact, the width of the peak is narrower, with a FWHM of 21\,keV. 

The peak energies at the positive- and negative-radii scanning positions have a discrepancy of up to 11\%. The bulk event energy scale, as measured with dedicated calibration runs and environment background $\gamma$ peaks, was stable to within 1\,keV (see \ref{app:calibration} for more details). This apparent asymmetry can be ascribed to the instability of the DCR parameter discussed at length in \ref{app:stability}. Subsequent measurements of the same scanning position (with the source unmoved between the measurements) demonstrate an average $\alpha$ energy peak drift of $0.2\%$ per day, to increasing energies. The ordering of the measurements matches the deviation seen in Fig.~\ref{fig:EvR}: measurements began with small-magnitude negative radii and moved to ever-larger-magnitude negative radii, followed by measurements moving from large-magnitude positive radii to small-magnitude positive radii. Scans of positive and negative small-radius positions were separated in time by up to three months; these positions also show the largest observed energy deviation. Minimal apparent asymmetry is seen in scans of large-radius positions, which occurred within one month of one another. 

\subsection{\label{ssec:dcr}\texorpdfstring{$\alpha$}{Alpha} Event Pulse Shape}
Depending on the charge collection properties of the detector at their point of incidence, $\alpha$ events can exhibit distinctive pulse shapes in a variety of ways. Two complementary discriminators, A/E and DCR, were studied in the TUBE scans. 

The A/E values of pulses in \ppc\ detectors depend strong\-ly on the event incidence radius \cite{GERDA_ae}, rising as the radius falls to 0. In detectors with small passivated surface radii, $\alpha$ events therefore have anomalously high A/E values, a signature used by the GERDA Collaboration to reduce the impact of $\alpha$ backgrounds \cite{GERDA2017}. In these measurements, the DCR effect reliably identifies $\alpha$ events at large radii, where A/E loses sensitivity as an $\alpha$ event discriminator. We study the value of DCR as a function of position on the passivated surface to determine how the expected $\alpha$ event rejection efficiency may vary depending on the geometric distribution of $\alpha$-emitting contaminants and to provide a detailed comparison to rising-edge-based $\alpha$ event discriminators. These measurements also allow us to study the mechanism underlying the DCR effect, as discussed in Sec.~\ref{sec:siggen}. 

\begin{figure}
  \centering
 \includegraphics[width=\linewidth]{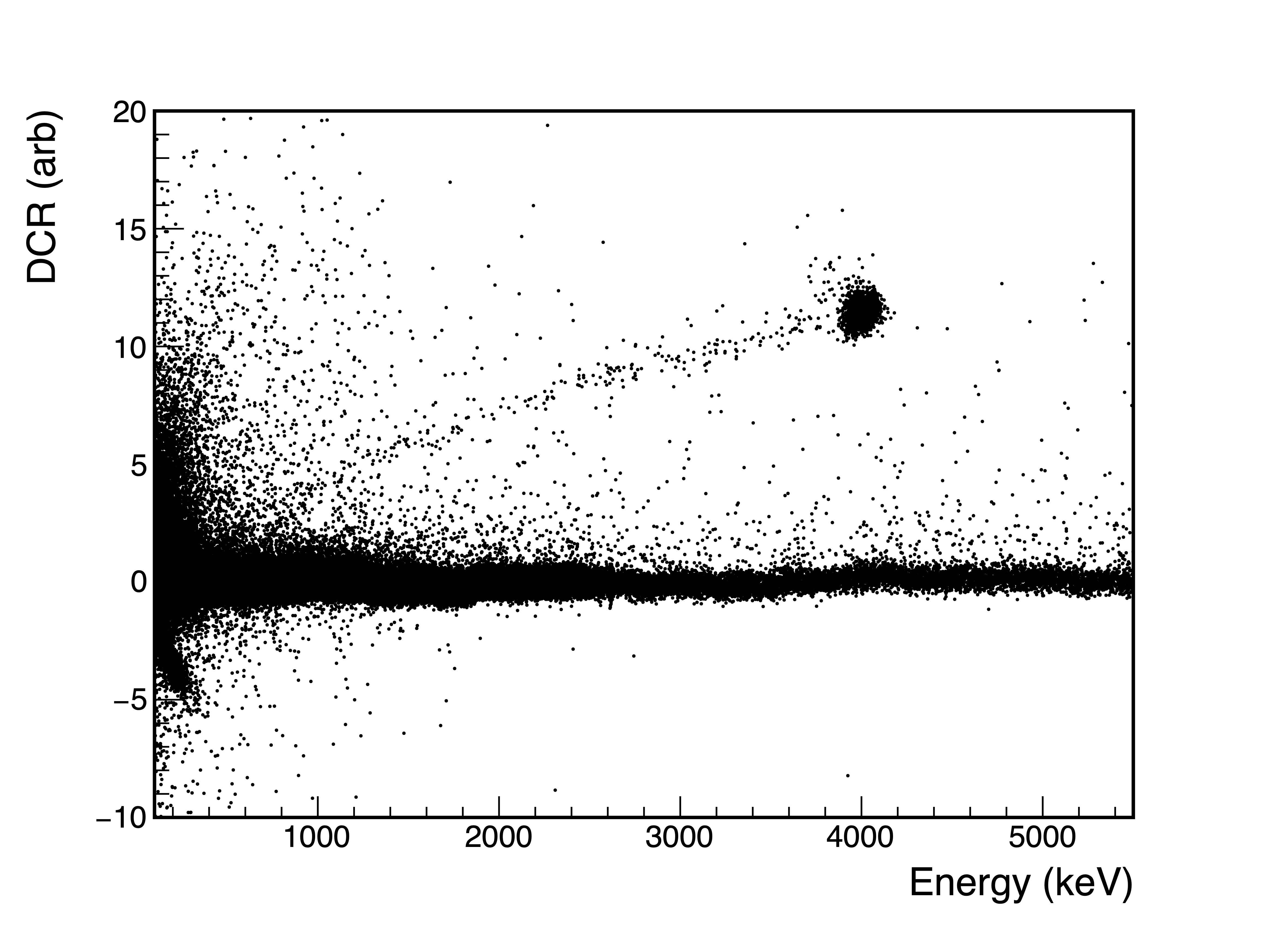}
 \caption{The DCR and energy distribution measured with the $\alpha$ source incident at $r = 18.0$\,mm. The tail of events degraded in energy and DCR is thought to be due to $\alpha$'s that deposit their energy at shallower depths, as discussed in Sec.~\ref{ssec:dcr}. Events with low values of DCR are caused by pile-up.}
 \label{fig:DCRvE}
\end{figure}

In the DCR distribution for each data set with $|r| \geq 4.5$\,mm, approximately 90\% of $\alpha$ events fall in a Gaussian peak. The remaining 10\% of events have both degraded energies and DCR values, as seen in Fig.~\ref{fig:DCRvE}. Follow-up investigations in the CAGE (Collimated Alphas, Gammas, and Electrons) scanner, which allows for measurements with a wide range of $\alpha$ beam angles of incidence, indicate that these events likely originate from shallower-depth energy depositions \cite{CAGEposter} \footnote{G. Othman, private communication.}. We find, however, that their DCR value and prompt energy have the same ratio as those events falling in the energy peak, indicating that the mechanism causing their DCR effect is consistent with the mechanism of slow charge release for events in the peak. To simplify the modeling of $\alpha$ events, we disregard the events falling outside of the energy peak when investigating the DCR response.

The DCR distribution for each scan is fit with a Gaussian, and the underlying background events are fit with a step function centered at the mean of the Gaussian. In the narrow fitting range used, the step function accounts for the high-DCR tail of the bulk event distribution. As in the analysis of the $\alpha$ peak energy, an additional A/E cut is applied to select near point-contact events when studying small-radius scans. Since the $\alpha$ events are better-separated from bulk events in DCR than in energy, this additional cut is only used for data sets with $|r|<3$\,mm when studying the DCR distribution. Events incident on the point contact itself do not have a distinct peak in DCR. In this case, the peak position is fit using an energy window in which the $\alpha$ events dominate the spectrum; a 5$\sigma$ window around the peak energy is used. See \ref{app:dcr} for details. 

\begin{figure}
  \centering
 \includegraphics[width=\linewidth]{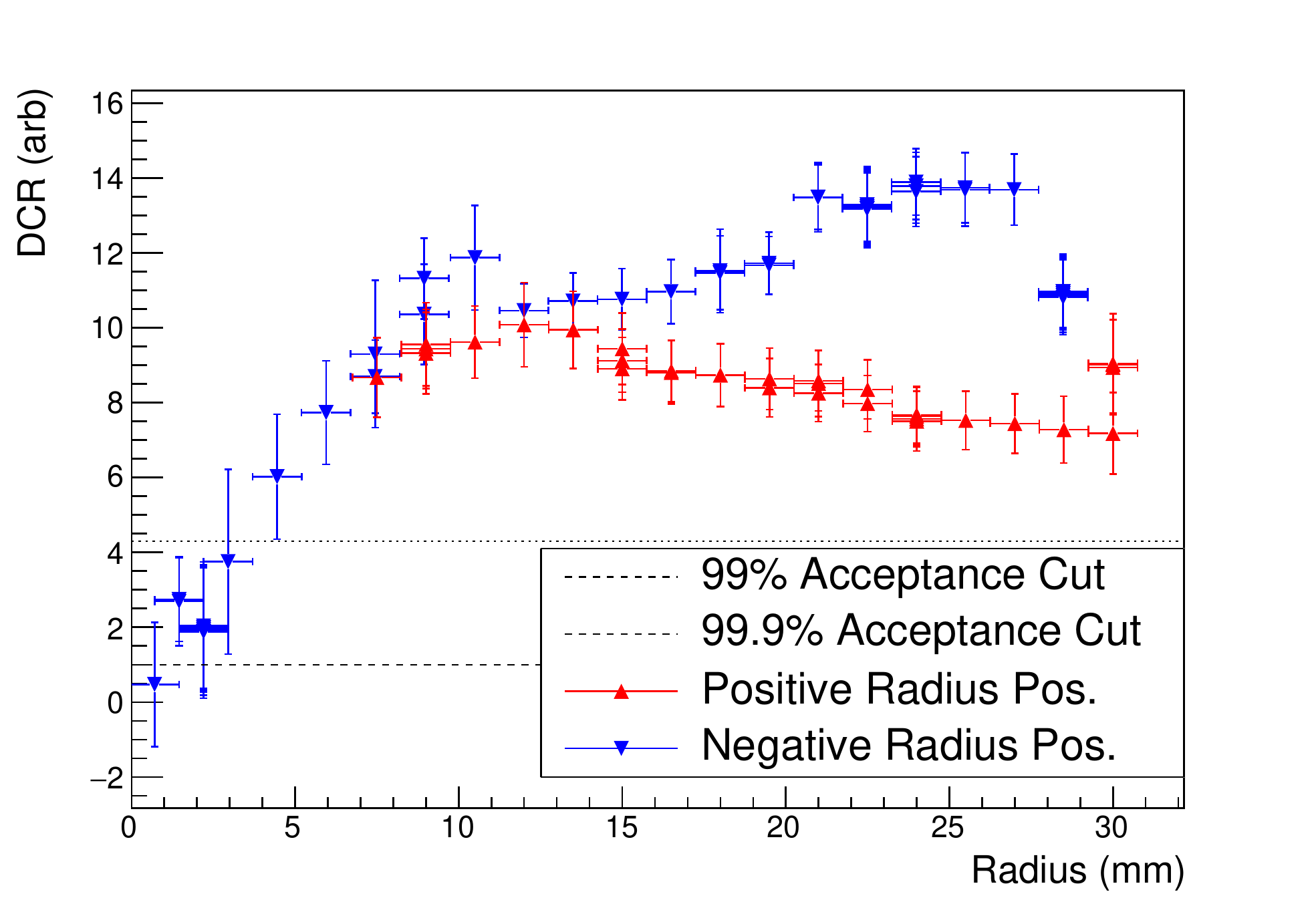}
 \caption{The centroids of the DCR peaks in each data set; certain positions were studied in multiple data sets. Negative (positive) radius positions are indicated by blue downward-pointing (red upward-pointing) triangles. The vertical error bars give the $5\sigma$-width of the DCR distribution peaks, and the horizontal errors depict the 0.75\,mm estimated uncertainty of the source position. The dashed (dotted) line indicates the 99\% (99.9\%) bulk event acceptance DCR cut. The observed instability of the DCR peak position is discussed in \ref{app:stability}.}
 \label{fig:DCRvR}
\end{figure}

\begin{figure*}
  \centering
 \includegraphics[width=\linewidth]{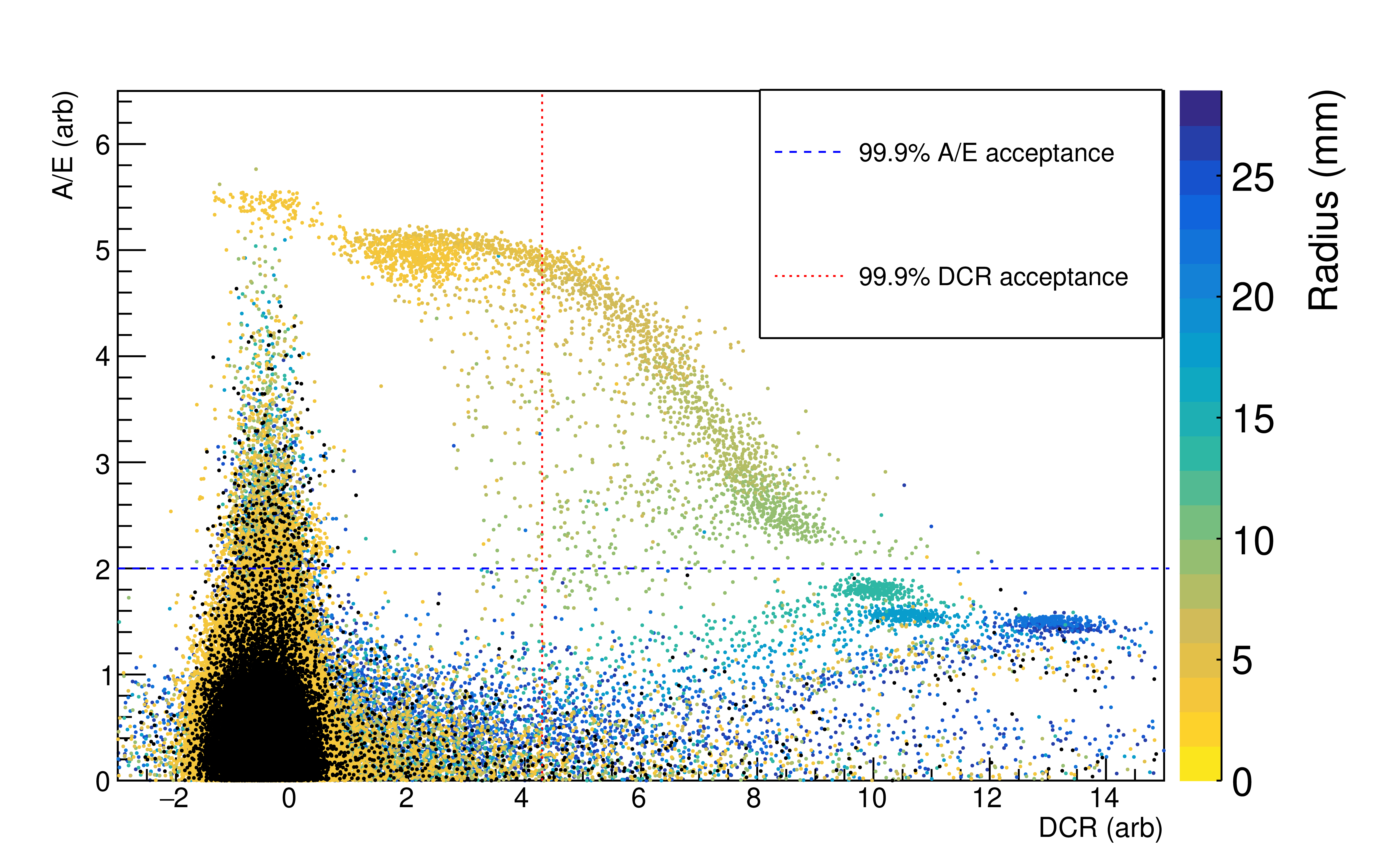}
 \caption{The distribution of A/E and DCR values for a range of scanning positions (indicated by the color scale) and a data set with no $\alpha$ source shining on the detector surface (in black). All single-site non-muon events with energies between 1 and 6\,MeV are included. The dashed blue line indicates the 99.9\% acceptance value of A/E and the dotted red line indicates the 99.9\% acceptance value of DCR.}
 \label{fig:AEvDCR}
\end{figure*}

The DCR peak positions found in the fits are depicted in Fig.~\ref{fig:DCRvR}, along with their $5\sigma$ peak widths. At positions with $|r|\geq4.5$\,mm, the DCR values of $\alpha$ events are consistently much higher than those of background events, and the DCR parameter is a highly efficient pulse shape discriminator, as shown in Sec.~\ref{ssec:eff}. 

The DCR values differ by up to a factor of 2 at the $0\degree$ and $180\degree$ scanning positions due to instability of the DCR parameter over the course of scanning. As discussed in \ref{app:stability}, several indicators of the passivated surface conditions point to positive charge build-up over time, driven by the incomplete $\alpha$ event charge collection. This instability had a minor but measurable effect on the observed energy of the $\alpha$ events, and a major effect on the observed slow charge recovery rate (i.e.~DCR). The instability led to rising values of DCR over time, which would improve the surface $\alpha$ rejection efficiency. The surface charge instability did not affect the bulk event charge collection efficiency or energy resolution, and is expected to be minimal in low $\alpha$ rate environments like those found in \nonubb\ searches. 

To further study the charge collection properties at the passivated surface, we perform exponential fits to the tails of a set of pole-zero corrected waveforms from each position included in the model fits. We limit the data used to the first 3000 runs taken (1500 hours, taken over the first 78 days of the first 217-day detector deployment), to reduce the effects of the surface charge instability. The average time-constant of delayed charge release, regardless of the position on the passivated surface, is found to be $\tau=900\pm100\,\mu$s. If the rate of delayed charge release remains constant over the full course of the event, this implies that time required to achieve full charge collection of an $\alpha$ event is approximately $5\tau=4.5$\,ms. 

Since the magnitude of the DCR effect observed in this detector is largest for events with large incident radii on the passivated surface, it is highly complementary to the A/E-based $\alpha$ rejection approach. The distributions of $\alpha$ events incident at various radii in the A/E vs. DCR parameter space (see Fig.~\ref{fig:AEvDCR}) clearly show the expected complementarity. At the smallest radii, the DCR effect is small, and the $\alpha$ events are not well-separated from bulk events in the DCR parameter space. These events, however, have larger A/E than 99.9\% of bulk events. At larger radii, the reverse is true; the $\alpha$ events have large DCR values and A/E values similar to those of many more bulk events. This suggests that the combination of the two parameters should provide effective $\alpha$ background rejection with the highest-possible bulk event efficiency.   

\subsection{\texorpdfstring{$\alpha$}{Alpha} Survival Probability}\label{ssec:eff}

\begin{figure}
  \centering
 \includegraphics[trim={0 2.7in 0 2.8in},clip, width=\linewidth]{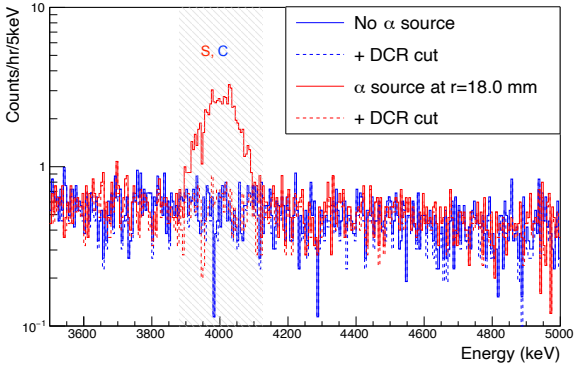}
 \caption{The energy spectra measured with the TUBE detector without the $\alpha$ source incident on the passivated surface, in blue, and with the source incident at r=18.0\,mm, in red. Spectra are shown before (the solid lines) and after (the dashed lines) the application of the DCR cut. The hatched energy window is used for the $\alpha$ survival probability calculation. The regions indicated here correspond to the variables in Eqn.~\ref{eqn:eff}.}
 \label{fig:eff}
\end{figure}

Though the expected \nonubb\ efficiency of the DCR pulse-shape discriminator can be determined using readily-available $\gamma$ calibration data (see Ref.~\cite{MJD2019}), the rate at which $\alpha$ background events are misidentified as signal-like can only be determined from this dedicated $\alpha$ event study. It is determined here as a function of radial position on the passivated surface using both the DCR-only $\alpha$ rejection approach and a combined DCR and A/E-based rejection approach. 

This quantitative analysis of the $\alpha$ event rejection focuses only on the full-depth events in the Gaussian event energy peak; the $\alpha$ survival probability for events in the tail of shallower-depth events is not calculated in this study. The fixed incidence angle of the source leads to a low rate for these events that varies dramatically with the radius of the scanning position. In preliminary simulations, the variation in the rate of this population based on changes to the scanning incidence angle is degenerate with variations due to conditions on the passivated surface (see Section~\ref{sec:siggen}). Therefore, any efficiency calculated for these events would be both high in uncertainty and of limited applicability to other detectors in other settings. The consistent ratio of the DCR to the prompt energy found in these events and the events in the peak leads us to expect that the $\alpha$ rejection will perform similarly well for tail events in the \nonubb\ ROI; the integrated energy of the delayed charge is 2-3\% of the total prompt signal, corresponding to 40 to 60\,keV at $Q_{\beta\beta}$. This effect is well above the energy resolution limit of HPGe detectors in \nonubb\ searches. This can also be seen from the $\alpha$ band separation from the bulk events at the Q$_{\beta\beta}$ energy in Fig.~\ref{fig:DCRvE}. A quantitative study of the rejection efficiency for these events is planned as part of the CAGE measurements \cite{CAGEposter}.

To calculate an $\alpha$ survival probability, we calculate the excess counts in the 5$\sigma$ region centered at the mean value of the $\alpha$ peak energy before and after applying a cut rejecting $\alpha$ events. The expected background rate is determined from the spectral shape measured in a source-free data set. The estimation window determination is described in~\ref{app:efficiency}. $S$ and $C$ are the signal regions in the data set taken with and without the source, respectively. The spectra before and after the application of the $\alpha$ event rejection cut are indicated by the subscripts $u$ and $c$, respectively. See Fig.~\ref{fig:eff} for an example. The $\alpha$ rejection efficiency is given by:
\begin{equation}
\epsilon = \frac{S_c - \tau C_c}{S_u - \tau C_u},
\label{eqn:eff}
\end{equation}
where $\tau$ is the ratio of the live time in the source run to the live time in the source-free run.

Event counts ($S_u$, $S_c$, etc.) are determined separately in each data set,  using the $\alpha$ signal energy windows and pulse shape parameters found for that data set. Then all Deployment 1 data sets taken at each particular source incidence position are combined to give a single efficiency evaluation for each position on the detector surface. The measurements in different regions are independent of one another, and given the low statistics, uncertainties on the count rates are assumed to be Poisson-distributed. 90\% confidence intervals for the efficiencies are calculated via ROOT's (see Ref.~\cite{root}) \texttt{TRolke} class with the profile likelihood method using the fully frequentist treatment described in Ref~\cite{Rolke}.  Where the $\alpha$ survival probability is found to be 0 or 1, an upper or lower limit is given.

Due to a slight (1 to 2\degree) misalignment of the scanning and IR cup axes, the source beam is partially occluded at negative radius scanning positions. This leads to reduced $\alpha$ event rates and increased statistical uncertainty at these positions; for some positions, these effects are mitigated by combining multiple data sets, leading to longer run times and higher statistics. For more discussion of factors affecting the uncertainty, see~\ref{app:efficiency}.

\begin{figure}
  \centering
 \includegraphics[width=\linewidth]{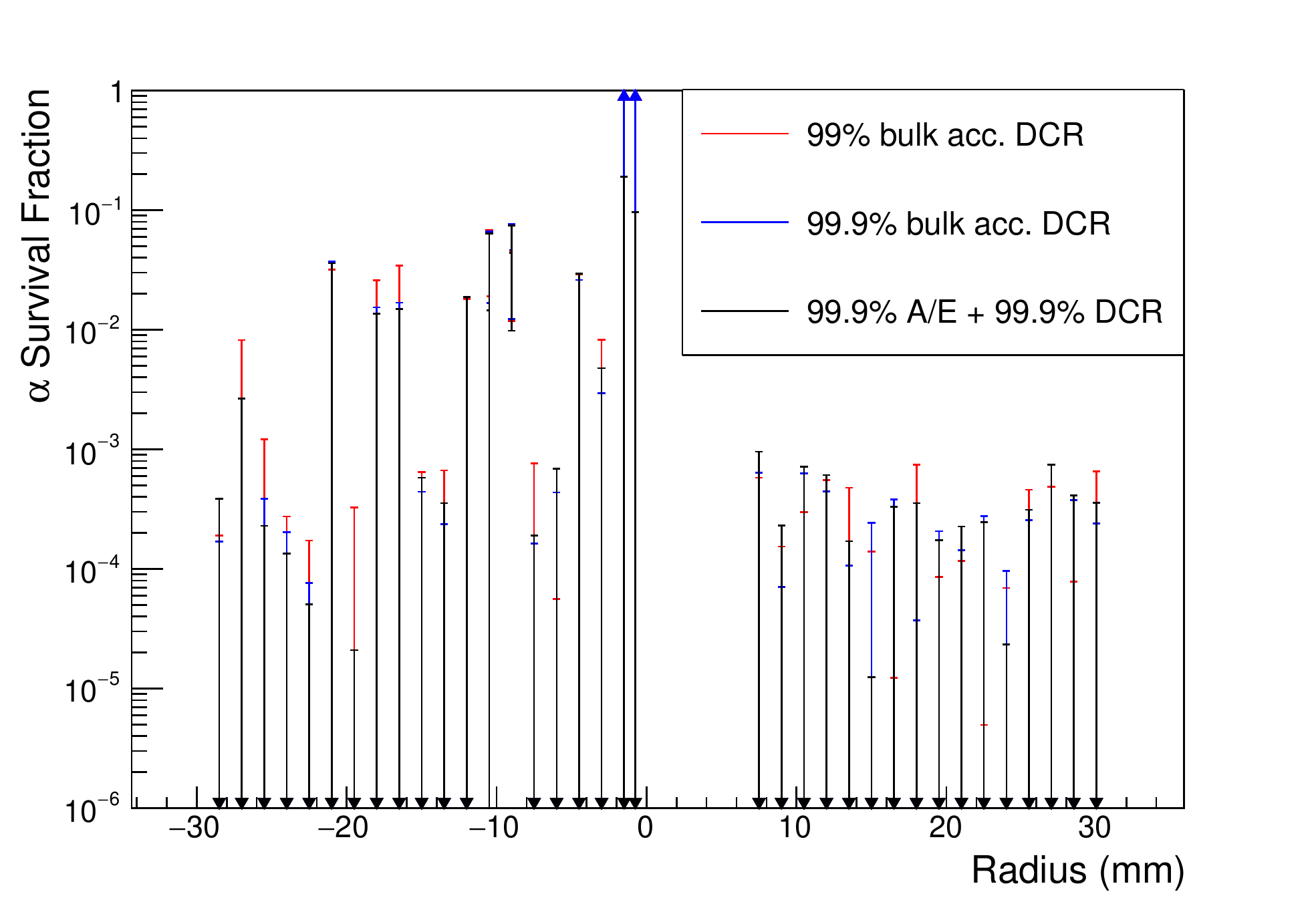}
 \caption{The $\alpha$ survival probability of three possible $\alpha$ event cuts, applied at each scanning position. The efficiency of the 99\% (99.9\%) bulk event acceptance DCR cut is given by the red (blue) points. The black points indicate the efficiency of an $\alpha$ event rejection cut using both the DCR and A/E parameters, each applied at the 99.9\% bulk event acceptance level. Downward-pointing arrows indicate that the uncertainty given is an upper limit.}
 \label{fig:rejEff}
\end{figure}

\begin{table}
\centering
\begin{tabular}{p{4cm} l l }
$\alpha$ Event Cut & Radius Range & $\alpha$ Survival (\%)  \\
\hline
99.9\% bulk acc. DCR & $|r|\geq 3$\,mm &  $<0.8$ \\
99.9\% bulk acc. DCR & $|r| < 3$\,mm &  $>8$ \\
99\% bulk acc. DCR & $|r|\geq 3$\,mm  & $<0.9$  \\
99\% bulk acc. DCR & $|r| < 3$\,mm &  $>0$ \\
99.9\% A/E and 99.9\% DCR & $|r|\geq 3$\,mm & $<0.8$ \\
99.9\% A/E and 99.9\% DCR & $|r|< 3$\,mm & $<14$ \\
\end{tabular}
\caption{The average $\alpha$ survival probability of each cut, found based on the 90\% confidence intervals. The average of the upper or lower limits is given, depending on the scanning position range in question. All included positions are weighted equally in the averages.}
 \label{tab:eff}
\end{table}
The $\alpha$ survival rates are evaluated with three different $\alpha$ event cuts: the DCR cut is tested at bulk acceptance rates of 99 and 99.9\%, and a cut using both DCR and A/E pulse shape parameters is also studied. In the latter case, we apply both a 99.9\% bulk acceptance DCR cut and a 99.9\% bulk acceptance A/E cut eliminating near-p$^+$ contact events, for a total bulk acceptance of 99.8\%.

The $\alpha$ survival probability for each cut as a function of passivated surface position is shown in Fig.~\ref{fig:rejEff} with upper and lower limits indicated by the arrows. The uncertainties shown are purely statistical. The average survival probability for each cut is also given in Table~\ref{tab:eff}: we consider the positions with $|r|\geq 3$\,mm separately from those with $|r|<3$\,mm, since the former have a survival probability central value at or very near 0, and the latter at or very near 1 for cuts using only DCR. Combining these ranges obscures the behavior of the $\alpha$ event cut. Any weighting strategy that attempts to account for the true incident $\alpha$ event rate at different positions on the detector surface requires an assumption about the $\alpha$ event source distribution and the effect of incidence angle. Since the incidence angle effect was not studied in these measurements and the source distribution depends on the experimental setup, all included positions are weighted equally in the average. We average the upper limits or lower limits of the survival probability, as is appropriate for each region. In one case ($r <3$\,mm with the 99\% acceptance DCR cut), the 90\% confidence interval on the survival probability is large enough to include the entire range from 0 to 100\%.

The $\alpha$ survival probability in every scan with $|r|\geq3$\,mm is consistent with 0\%, within 90\% confidence, and all scanning positions have a survival probability upper limit of less than 7.5\%.  Though the 99\% bulk acceptance cut does remove more events from each data set than the looser 99.9\% bulk acceptance cut, the upper limits found are dominated by the uncertainty due to fluctuations in the background level, leading to a slightly lower upper limit in the looser cut. At the positions closest to the p$^+$ contact, only the combined DCR and A/E-based $\alpha$ identification parameter is effective. Even in the relatively high-noise environment of the TUBE cryostat, this combination of pulse shape discriminators eliminates $\alpha$ background events, with a maximum survival probability upper limit of 19\%, across the entire detector surface with just 0.2\% signal event loss. 

\section{Modeling the \texorpdfstring{$\alpha$}{Alpha} Event Response}\label{sec:siggen}
Using simulations of signal formation in the detector, we can investigate the mechanism underlying the observed $\alpha$ energy degradation and delayed charge effect. We first model the potential causes of energy loss to distinguish which charge carrier(s) are responsible for the observed energy degradation. Based on our findings, we then consider which of the affected carriers could be responsible for the observed delayed charge. The two sets of models considered (for energy and DCR) are therefore interrelated but not identical.

The detector may have a true, fully inactive layer on the passivated surface. Such a dead layer would affect electrons and electron-holes in exactly the same way, and the $\alpha$ energy loss incurred would not vary by position on the passivated surface.  The maximum allowable thickness of this dead layer can be calculated from the maximum observed energy of $\alpha$s on the passivated surface, which is 4.746\,MeV. Using the Bethe formula as calculated in the NIST ASTAR database \cite{SRIM}, we find that the maximum dead layer that is compatible with our observations is 3.5\,$\mu$m thick, far thicker than anticipated based on our understanding of the passivation process.

In TUBE, the observed energy of $\alpha$ interactions varies dramatically depending on the incident radius of the $\alpha$. This indicates that while a thin dead layer may be present, charge loss must also be occurring. The radial dependence of energy indicates that positive and negative charge carrier contributions are affected differently. Therefore, we use a more sophisticated model of signal formation to investigate the detector response to surface $\alpha$ events. 

\begin{figure}
  \centering
 \includegraphics[trim={0 0.5in 0 0.22in},clip, width=\linewidth]{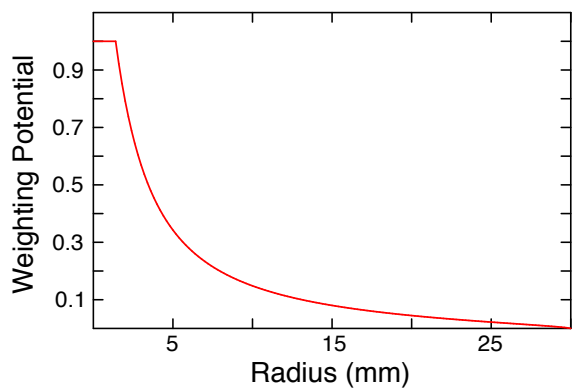}
 \caption{The weighting potential at $z=0$ (i.e. events originating at the passivated surface) for the PONaMa-1 detector, calculated using {\tt mjd\_fieldgen}. Note that the electron contribution to the signal is small over most of the surface, and rises dramatically at small radii.}
 \label{fig:wp}
\end{figure}

As described in Ref.~\cite{He2000}, the Shockley-Ramo theorem can be used to calculate the induced signal expected at the point-contact of the detector. It can be written in terms of the weighting potential $\Phi(x)$ to give the integrated charge at the point contact:
$$
Q(t) = q\Delta\Phi(x(t))
$$
where $q$ is the charge in motion in the detector and $Q(t)$ is the charge induced at the point contact. The value of the weighting potential at a point can be conceptualized as the fraction of the total induced charge resulting from the minority carrier (in this case, electron) motion over the course of the full signal evolution. The value of the weighting potential varies as a function of the initial interaction position within the detector, and can be used to decompose the signal into the portions induced by the positive and negative charges. The weighting potential of the detector studied here is shown in Figure~\ref{fig:wp} for $z = 0$, describing interaction positions at the passivated surface.

Signals are simulated using the {\tt mjd\_siggen} software package\,\cite{siggen}, which has two parts. From the known impurity gradient and geometry of the detector, the electric field inside the detector is calculated using the stand-alone {\tt mjd\_fieldgen} software. The {\tt mjd\_siggen} software is then used to calculate the expected total energy as a function of radius for events occurring at the passivated surface of this detector. More details on the pulse shape simulation of this detector can be found in Ref.~\cite{Mertens_siggen}. 

We limit the data used for fits to {\tt siggen}-based models of charge collection on the passivated surface to just the first 3000 runs (1500 hours, taken over 78 days) of the first 217-day detector deployment period. Using the data from this period allows us to reduce the effect of the surface charge instability while retaining measurements from across a full range of radial source positions. The energy and DCR peak positions found from fits of the 42 included data sets are averaged together by position, with fit uncertainties summed in quadrature. Datasets with $r<3$\,mm are excluded from the fits, since at these scanning positions, the source is partially or fully incident on the p$^+$ contact. 

\begin{figure}
\centering
  \subfloat[][Three different models of $\alpha$ energy loss are shown along with their dead-layer variants: Model 1 (1d) is indicated by the dashed (dotted) line, the results of Model 4 (4d) are given by the red crosses (asterisks), and the results of Model 3 (3d) are given by the violet triangles (`X'es). Model 2 is not shown because its best-fit is found when the electron component contribution is 0\%, making it identical to Model 3.]{
   \includegraphics[width=\linewidth]{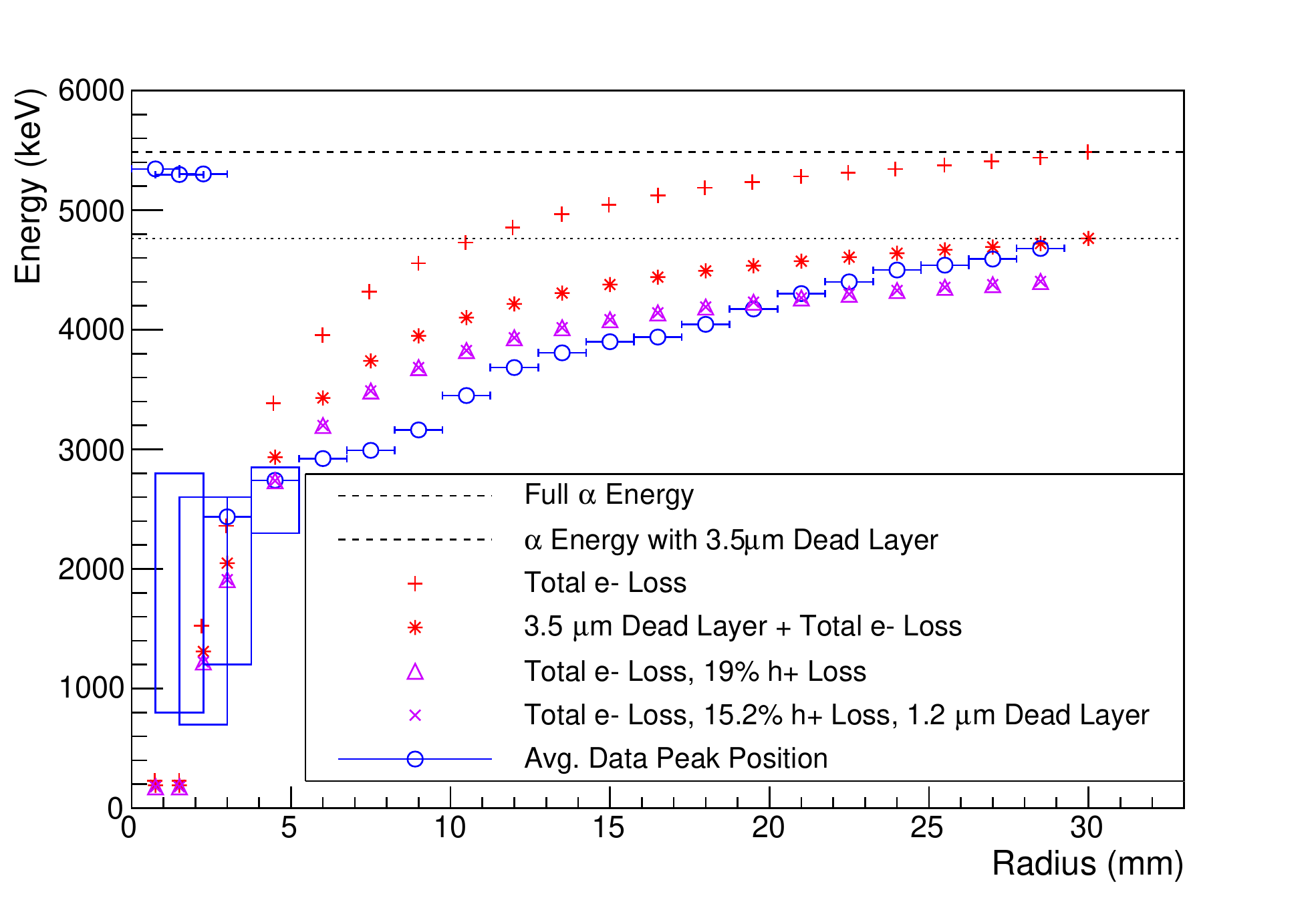}
  \label{fig:siggen}}
   \hspace{3mm}
    \centering
 \subfloat[][Two different models of $\alpha$ delayed charge recovery are shown: the results of Model B are given by the red crosses, and the results of Model C are given by the violet `X'es. Model A is not shown because its best-fit is found when the electron component contribution is 0\%, making it identical to model C.
 ]{
 \includegraphics[width=\linewidth]{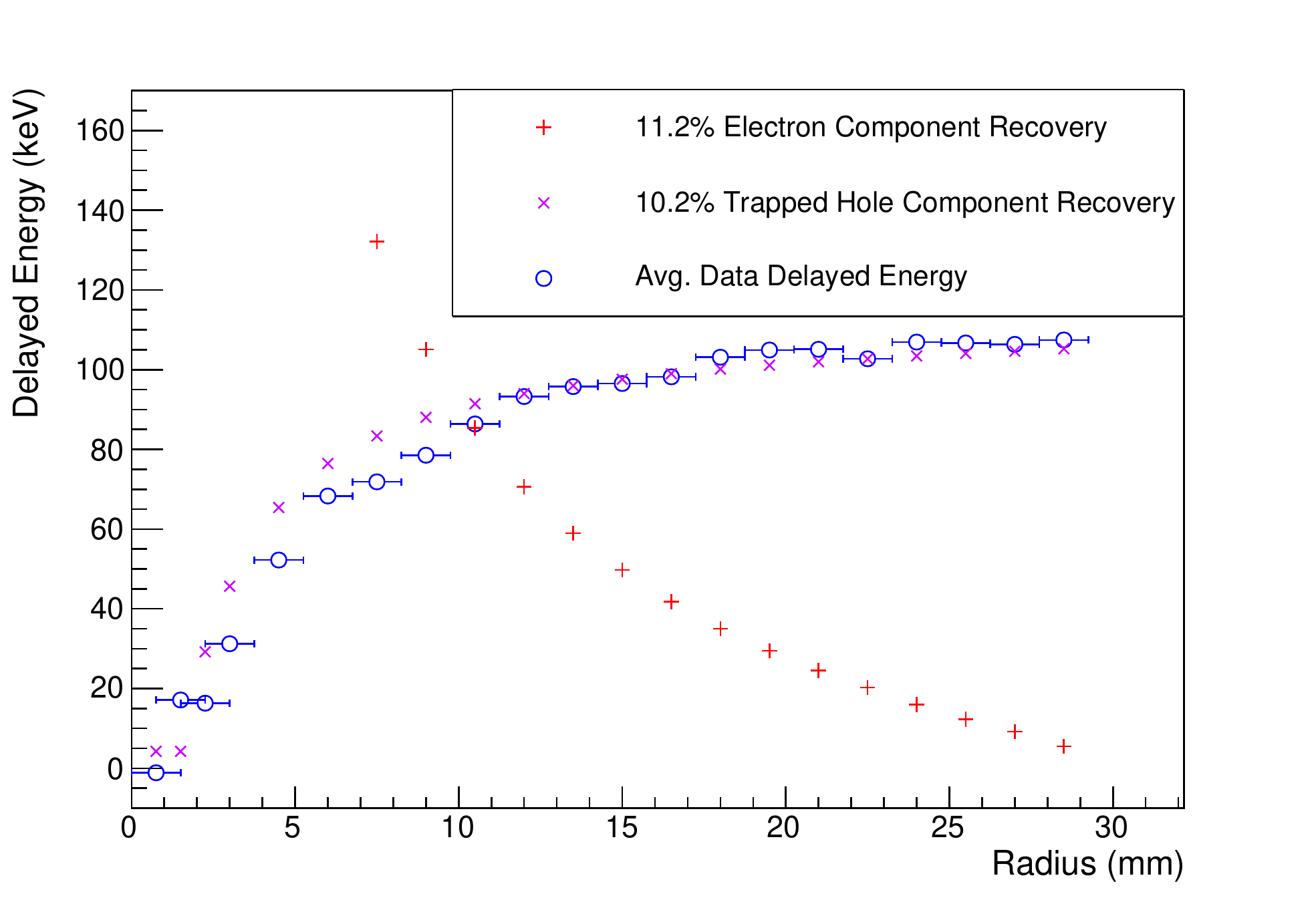}
 \label{fig:siggen_DCR}}
\caption{The results of {\tt siggen} simulations of energy loss \textit{(a)} and delayed charge recovery \textit{(b)} of events occurring on the passivated surface. The average values of spectral/DCR distribution fits to the data are given by blue circles, with blue boxes indicating the $\alpha$ event energy range in cases where the peak position does not adequately capture the distribution shape.}
\end{figure}

The {\tt siggen} charge collection behavior can be modified to reflect various charge loss mechanisms. Several models are considered: 
\begin{itemize}
\item Energy Model 1: Charge collection for both electrons and electron-holes occurs as it does for bulk events. This model is provided for reference.
\item Energy Model 2: Some fraction of each of the electron and electron-hole signals does not contribute to the prompt energy. The fractions are held constant with respect to radius but allowed to float independently in the fit. This model matches the behavior expected if near-surface trapping and/or slow surface drift of both electrons and electron-holes were contributing to the prompt energy loss, with some measurable fraction of the electron component collected promptly.
\item Energy Model 3: The electron component does not contribute to the prompt energy, and a fraction of electron-holes (held constant with respect to radius, but allowed to float in the fit) is trapped or slowed instead of being collected promptly. This model matches the behavior expected if near-surface trapping and/or slow surface drift of both electrons and electron-holes were contributing to the prompt energy loss, with none of the electron component collected promptly.
\item Energy Model 4: The electron component does not contribute to the prompt energy, and electron-holes are collected as they are in bulk events. This model matches the behavior expected if electron surface drift or trapping were the only source of prompt energy loss.
\end{itemize}

Models 2-4 are all potential behaviors exhibited by a detector in which positive charges are present on the passivated surface , with Model 2 being most likely at low surface charge densities, where self-repulsion of the charge cloud dominates the $\alpha$ energy loss. Models 3 and 4 reflect the expected behavior at higher positive charge densities, where electrons are effectively pulled to the passivated surface. The instability in the charge drift time seen in these measurements (see \ref{app:stability}) suggests that positive charges may be accumulating on the surface over the course of the scanning measurement. 

Another category of energy loss models, corresponding to Models 2-4 but with the signs of the charges reversed, is also possible. In these models, electron-holes exhibit surface drift or near-surface trapping, and are partially or entirely missing from the prompt signal. These models, which describe possible behaviors of a detector with a net-negative charge on its passivated surface, lead to $\alpha$ energies that fall with increasing incidence radius. Since this is opposite to the behavior we observe, we did not study these models in detail. Under different operating conditions, however, negative charge build-up could occur, and these models would be appropriate descriptions of the observed behavior, as discussed in Ref. \cite{GALATEA_scan}.   

We also study a variation of Models 1-4 that allows for the possibility of a fixed-thickness dead layer over the entire passivated surface. Such a dead layer would degrade incoming $\alpha$ particle energies by a fixed amount, regardless of their position. In the case of Models 1 and 4, the dead layer thickness is set to its maximum allowed value of 3.5\,$\mu$m. In the case of Models 2 and 3, it is allowed to float. Table~\ref{tab:E_models} summarizes which parameters are held fixed and which are allowed to float in each model's fit. The variant of each model with an added dead layer thickness is indicated by the addition of a `d' to the model number.

Using only the centroid values from the passivated surface $\alpha$ peaks, we minimize the residual sum of squares between the model and the peak data. Since the systematic uncertainty associated with the charge collection instability is not well-characterized and we cannot assume that this uncertainty is Gaussian-distributed, the absolute value of the reduced $\chi_{red}^2$ (i.e.~whether its value is near 1) cannot be used to quantify the goodness-of-fit. We can, however, use it for model comparison, and to find the best-fit value for the electron-hole and electron contributions in Models 2 and 3. 

\begin{table}
\centering
\begin{tabular}{l l l}
E Model & Free Param. &  Fixed Param.\\
\hline
1 & None & $f_e = 1, f_h = 1, d = 0\,\mu$m \\
1d & None & $f_e = 1, f_h = 1, d = 3.5\,\mu$m \\
2 & $f_h, f_e$ & $d = 0\,\mu$m \\
2d & $f_h, f_e, d$ & None \\
3 & $f_h$ & $f_e = 0, d = 0\,\mu$m \\
3d & $f_h, d$ & $f_e = 0$ \\
4 & None & $f_h = 1, f_e = 0, d = 0\,\mu$m \\
4d & None & $f_h = 1, f_e = 0, d = 3.5\,\mu$m \\

\end{tabular}
\caption{A summary of the $\alpha$ energy response models studied. $f_e$ and $f_h$ are the fraction of the electron and electron-hole charge collection signals that contribute to the prompt event energy, and $d$ is the dead layer thickness.}
 \label{tab:E_models}
\end{table}

Models 1, 3, 4, and their dead-layer variants are shown in Fig.~\ref{fig:siggen} along with the average peak positions in the data and the $\alpha$ energy ranges observed in the smallest-radius data sets. The best-fit is found in Model 3, when 19\% of the electron-holes are lost to charge trapping or slow surface drift. 

The addition of a dead layer does not improve the good\-ness-of-fit, but it does result in a non-zero best-fit thickness. In this case, the best fit is found when incident $\alpha$ particles lose 245\,keV upon entry, corresponding to a dead layer that is 1.2\,$\mu$m thick, and 15\% of the electron-holes are lost to charge trapping. Though a thin dead layer may be present, these measurements are not able to distinguish its effects from those of incomplete electron-hole component collection. Future measurements of the passivated surface with lower-range incident particles, like low-energy $\gamma$ and $\beta$ sources, should allow us to study these effects independently. 

The preferred fit in Model 2 was found when the electron component of the signal was 0\%. We can conclude that if electrons do contribute to the measured prompt energy, their effect is below the sensitivity threshold of these measurements. Determining this contribution would require measurements to be conducted using a setup with a smaller source spot size on the detector surface. This would allow a finer-grained investigation of $\alpha$'s incident at small-radius positions, where the weighting potential is both high and rapidly varying.

Model 3 matches the observed $\alpha$ energy most closely, but does not match the data exactly. A possible source of the deviation is the instability, which is described in more detail in \ref{app:stability}. As discussed there, the prompt energy of the passivated surface $\alpha$ events increased over time. This corresponds to a decrease in the electron-hole loss fraction over time or an increase in the fraction of trapped electron-holes that are released into the bulk within the first 3\,$\mu$s of the signal. This is the behavior that would be observed if the passivated surface were becoming more positively charged over time. Variation in the trapping layer properties as a function of radius could also be responsible for some of the deviation. Both of these possibilities will be explored in future measurements. 

Given this near-match, we conclude that all or nearly all of the negative charge is being trapped and/or slowed for interactions near the passivated surface, and that a fraction of the positive charge is also being trapped and/or slowed. These behaviors occur for events on the passivated surface regardless of the radial position of the interaction. Events incident on the p$^+$ contact, on the other hand, do not show indications of significant charge trapping. The average energy loss observed at these positions is consistent with the loss seen in scans of the point contact of BEGe-type \ppc\ detectors \cite{AgostiniThesis}, and corresponds to a dead layer of 0.3 to 0.5\,$\mu$m.

Since we can conclude that both electrons and electron-holes are trapped and/or slowed in passivated surface events, the slow release of charges from either component could be responsible for the observed DCR effect. We consider the effects of both possibilities, and compare them to total energy collected as delayed charge at each scanning position. In this study, we disregard the effects of a thin fully-dead layer at the passivated surface, which would be minimal. The delayed energy is calculated as given in Eqn.~\ref{eqn:delayedE}, which gives the amount of energy collected as delayed charge in the first 18\,$\mu$s of waveform digitization following the end of the prompt rise. 

Three models of the DCR effect are considered:
\begin{itemize}
    \item DCR Model A: Both charge carrier species are allowed to contribute to the energy released as delayed charge. Each contributes a constant fraction of the component's total prompt energy, as described below. 
    \item DCR Model B: The DCR effect is caused by the electron component of the signal, whether from electron drift on the passivated surface or trapped charge re-release.
    \item DCR Model C: The DCR effect is caused by the electron-hole component of the signal, whether from electron-hole drift on the passivated surface or trapped charge re-release.
\end{itemize}
In Models B and C, the amount of delayed energy recovered in the first 18\,$\mu$s is a constant fraction of the total energy induced by the relevant charge carrier at each position. This is the case regardless of whether surface drift or near-surface trapping is responsible for the DCR: the surface charge drift velocity is constant for a particular drift direction relative to the crystal axis in the former case, and the charge trapping and subsequent re-release is assumed to affect a constant fraction of the charge carriers in the latter case.

As in the case of the energy models, these correspond to the potential behaviors exhibited by a detector in which a net-positive charge is on the passivated surface. A different set of models, with the signs of the charges reversed, would be relevant if the behavior of the $\alpha$ energies indicated that a net-negative charge was present on the passivated surface. 

As seen in Fig.~\ref{fig:siggen_DCR}, {\tt siggen} simulations of Models B and C show opposite behavior as a function of the radial position on the passivated surface. Table~\ref{tab:DCR_models} summarizes the free and fixed parameters in each model. The best-fit is found in Model C, when 10.2\% of the missing electron-hole component (i.e.~1.9\% of the total electron-hole component of the signal) is collected during waveform digitization. Adding the effect of a passivated surface dead layer increases these fractions slightly, but does not change the qualitative behavior. As in the energy degradation model, the electron component contribution to the delayed energy was driven to 0 in Model A. Slow electron transport or charge re-release may also be occurring, but it contributes negligibly to the signal shape in the 18\,$\mu$s window studied here. 

\begin{table}
\centering
\begin{tabular}{l l l }
DCR Model & Free Param. &  Fixed Param.\\
\hline
A & $g_e, g_h$ & None \\
B & $g_e$ & $g_h = 0$  \\
C & $g_h$ & $g_e = 0$  \\
\end{tabular}
\caption{A summary of the $\alpha$ delayed charge collection models studied. $g_e$ and $g_h$ are the fraction of the electron and electron-hole charge collection signals that contribute to the delayed event energy.}
 \label{tab:DCR_models}
\end{table}

Model C correctly predicts that low DCR values will be observed at the smallest radii. Though the value of DCR at each position changes over time in these measurements (as discussed in \ref{app:stability}), DCR remains lower at small radii than at large radii. This indicates that though the electron-hole trapping and/or release rate may be unstable, under the conditions found in the TUBE scanner, the electron-holes are always responsible for the DCR effect.

Though {\tt siggen}-based pulse shape simulations allow us to determine that both charge species are being lost or delayed, and that the electron-holes are responsible for the DCR effect, more detailed simulations that model the behavior of the full charge cloud are needed to determine whether surface charge drift effects are sufficient to explain the observed pulse shapes. These simulations are currently being developed, and will be discussed in a future publication.

\section{\texorpdfstring{$\alpha$}{Alpha} Backgrounds in \texorpdfstring{$^{76}Ge$}{Ge-76} \texorpdfstring{\nonubb}{0nBB} Searches }
The complementarity of rising-edge-based and de\-layed-charge-based $\alpha$ background identification strategies motivates the differing approaches taken by the \textsc{Majorana} and GERDA collaborations. For detectors in which the $\alpha$-sensitive surface is small in radius, like those used in the GERDA experiment, a drift-time-based pulse shape parameter like A/E can be used as the sole strategy to reject $\alpha$ events with high \nonubb\ signal efficiency. The enriched detectors used in the \MJ\ \DEM, however, have much larger-radius $\alpha$-sensitive surfaces; the \nonubb\ signal efficiency of such a cut would be unacceptably low. In this case, the DCR parameter is a far more effective alternative. 

In the \textsc{Gerda} Experiment's point-contact detectors, 234 $\alpha$ events at high energy are identified in 5.8\,kg\,yr of exposure \cite{GERDA2017}, indicating an $\alpha$ rate of at least 0.110\,counts/(kg\,day). Since Mirion's BEGe geometry \cite{Budjas2009} (which has only a small annular passivated ditch region) is used for the point-contact detectors, a cut on high-A/E events can serve as the sole identifier of $\alpha$ events \cite{GERDA2018}. This rising-edge-based cut incurs a fiducial volume loss of $(2.69\pm 0.06)$\% in the $^{228}$Th double-escape (DEP) peak, which is used as a proxy for \nonubb\ events \cite{WagnerThesis}. 

Given the larger-radius passivated region of the \MJ\ \DEM\ enriched detectors, a similar approach would incur a fiducial volume loss of over 10\%. The DCR method of $\alpha$ event identification, on the other hand, has a signal sacrifice of between 0.8 and 3.1\%, depending on the noise conditions of the data set \cite{MJD2019}. Using the DCR discriminator, an $\alpha$ rate of at least 0.276\,counts/(kg\,day) is identified between 1 and 5.5\,MeV. Unlike the high-A/E discriminator, this method is relatively insensitive to near-p$^+$ $\alpha$ interactions; if all the events between 2.7 and 5.5\,MeV that are retained after the application of all analysis cuts \cite{MJD2019} are assumed to be due to $\alpha$ interactions, the unidentified $\alpha$'s contribute an additional $7.5\times10^{-3}$\,counts/(kg\,day). Based on the results shown above (see Fig.~\ref{fig:AEvDCR}), the implementation of an additional rising-edge-based estimator tuned to sacrifice 0.1\% of bulk events would be expected to eliminate this remaining background contribution. Future analyses of the \MJ\ \DEM\ data will employ both DCR and high-A vs.~E discriminators to identify $\alpha$ events. The cut levels of both discriminators will be optimized based on \textit{in situ} measurements. 

\begin{figure}
  \centering
 \includegraphics[trim={0.5in 2.8in 0.7in 2.8in}, clip, width=\linewidth]{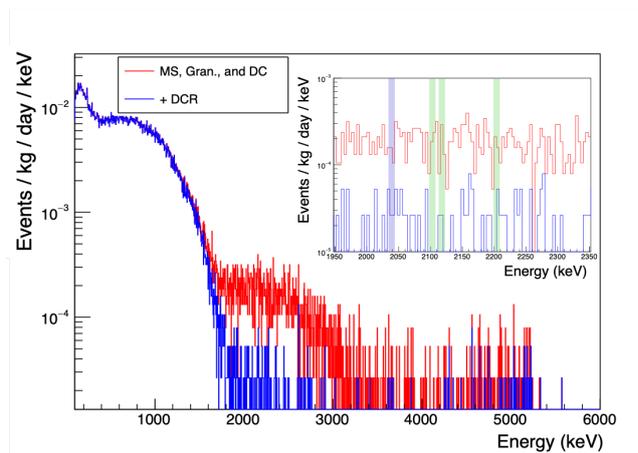}
 \caption{Energy spectra from the \MJ\ \DEM's enriched detectors, including all data from Datasets 0 through 6 \cite{MJD2019}. The red lines indicate the events kept by all other analyses (data cleaning, environmental, muon veto, multiplicity, and A vs.~E multi-site cuts). The blue line shows the events remaining after all cuts, including the DCR cut. The inset shows the effect of the DCR cut in the background estimation window. The regions excluded due to $\gamma$ backgrounds are shaded in green and the 10\,keV window centered on $Q_{\beta\beta}$ is shaded in blue.}
 \label{fig:MJD_spec}
\end{figure}

%However, the measurements described above indicate that these events are unlikely to lie in the \nonubb\ ROI, and should appear at higher energies unless they are significantly degraded before reaching the detector surface. 
The DCR cut is most effective at identifying highly-degraded passivated surface $\alpha$ events, and is less effective for events where a large fraction of the incident energy is collected promptly. This trend can be seen in Fig.~\ref{fig:MJD_spec}, where the survival fraction of events with energies above 2615\,keV increases with increasing energy. Given this behavior, $\alpha$ events that lie in the \nonubb\ ROI and survive the DCR cut would have to be significantly degraded in energy before reaching the detector, without additionally exhibiting delayed charge energy loss. Such events can still be identified using a high A/E or A vs.~E discriminator over much of the detector surface, and their rate  will drive the optimal cut levels determined for the DCR and rising-edge-based discriminators.

%which are most likely to impact the background rate in the ROI. These features are seen in the spectrum of events identified as $\alpha$ interactions by the DCR discriminator and in high-energy events remaining after all cuts (see Fig.~\ref{fig:MJD_spec}). 
The spectrum of identified $\alpha$ events shown in Fig.~\ref{fig:MJD_spec} also demonstrates structure from 1 to 3\,MeV, a feature that can be used along with charge transport and collection simulations of the passivated surface to constrain the spatial distribution of the $\alpha$ emitters when building the full \MJ\ \DEM\ background model.

Background projections based on the \MJ\ assay program and simulations predict a flat background between 1950 and 2350 keV after rejecting possible $\gamma$ peaks within that energy range. We exclude $\pm 5$\,keV ranges centered at 2103\,keV ($^{208}$Tl single escape peak), 2118 and 2204\,keV ($^{214}$Bi), and 2039\,keV (\nonubb) from the background estimation window. In the remaining 360\,keV window, the use of the DCR discriminator reduces the background rate  from $69.5\times10^{-3}$\,counts/(keV\,kg\,yr) to $6.1\times10^{-3}$\,counts/(keV\,kg\,yr), over an order of magnitude reduction \cite{MJD2019}. Suspected $\alpha$ events do survive at energies between 2.7 and 5\,MeV, indicating that some $\alpha$ events may also survive in the $0\nu\beta\beta$ region of interest. These events may be occurring on or near the $p^+$ contact, with the incoming $\alpha$ being degraded in energy due to interaction in inactive materials (i.e. in the source of the background events). These $\alpha$ events would be effectively suppressed by the implementation of a high-A vs.~E cut, which will be used in future \MJ\ analyses.

The upcoming LEGEND experiment \cite{LEGEND_pCDR} will combine techniques from both the GERDA and \MJ\ \DEM\ experiments, including the use of both rising-edge and DCR-based $\alpha$ rejection. LEGEND-200, the first stage of this project, will use existing \ppc\ detectors of varying geometries, as well as new inverted-coaxial detectors \cite{DOMULA2018106}. Our experience with PPC detectors demonstrates that the effect of a large passivated surface has to be taken into account when rejecting $\alpha$ events in analysis, and that the unique charge collection properties at this surface can help reduce the impact of such events in low-background experiments. 

We are also exploring the possibility of detector designs that minimize or avoid passivation entirely. Further study of these and existing BEGe-type detectors will focus on the effectiveness of the DCR technique in these alternative geometries. LEGEND, like GERDA, will use a liquid argon shield, potentially leading to $^{42}$K $\beta$ background events from the decay of $^{42}$Ar. Future studies of charge collection near the passivated surface will determine whether these events exhibit a DCR component, allowing them to be identified.

\section{\label{sec:conc}Conclusion}
This paper presented data taken with a p-type point contact detector produced by ORTEC that reproduces the design of the \MJ\ \DEM\ enriched detectors. The design of the TUBE cryostat and detector mount allows the passivated surface and p$^+$ contact of the detector to be scanned using a collimated $\alpha$ source. This enables a detailed study of the detector response to backgrounds from surface deposition of $^{222}$Rn progeny on parts and the detectors themselves. Due to degradation of the $\alpha$ particle energy in inactive materials and poor charge collection in the detector itself, these events can be a significant background source in experiments like the search for \nonubb\ in $^{76}$Ge. The work shown here, however, demonstrates that these events can be effectively rejected based on their pulse shape characteristics, while maintaining high signal event efficiency.

We find that $\alpha$'s incident on the p$^+$ contact do not exhibit the DCR effect and are observed at nearly their full energy. $\alpha$'s incident on the passivated surface, on the other hand, exhibit significant charge trapping and can be highly degraded in energy, depending on their radius of incidence. Modeling the process of signal-formation in this detector, we find that the observed $\alpha$ event energies are most consistent with a total loss of the electron-driven component of the signal, and a loss of between 15 and 19\% of the electron-hole-driven component. The energy recovered as delayed charge in the first 18\,$\mu$s following the prompt signal corresponds to a recovery of 10.2\% of that trapped electron-hole charge. The trapped electron-holes appear to be released with a 0.9\,ms time constant. The sharply reduced energies and DCR effects measured at these near-p$^+$ positions suggest that electrons do not play a significant role in the signal formation in the timescale of digitization. The formation mechanism of the DCR signal makes the DCR-based pulse shape discriminator highly effective at identifying $\alpha$ events on the passivated surface of the detector, particularly for $\alpha$'s incident at positions far from the p$^+$ contact. 

Subsequent measurements of this same detector in the $\alpha$ event characterization test-stand described in Ref.~\cite{GALATEA_scan} have found opposite behavior from that observed with the TUBE scanner, with the DCR component of $\alpha$ events decreasing with increasing event incidence radius. In this case, the DCR parameter can still be used to identify $\alpha$ events in the energy range relevant for \nonubb\ searches. These measurements seem to indicate that the passivated surface becomes negatively charged, instead of the positive charge build-up we observe. This indicates that environmental conditions may impact the observed $\alpha$ event response. In this case the vacuum conditions are believed to play a major role in the differing observed behavior. Additional measurements that study the effect of detector bias voltage, the position of the ground plane, the $\alpha$ incidence angle, and the detector temperature are underway.

Detailed pulse-shape simulations that correctly model the effects of self-repulsion and diffusion in the dense charge cloud created by $\alpha$ interactions have been developed in response to the behavior observed in this study. These simulations will be used to study the effect of varying surface charge densities and polarity on the pulse shape and energy observed in $\alpha$ interactions. Along with additional detector surface characterization measurements that are being performed as part of the LEGEND experiment, these studies will inform the $\alpha$ background model being developed for the \MJ\ \DEM.

Based on the presence of the DCR component and existing rising-edge-based $\alpha$ discrimination techniques, we are able to reliably identify $\alpha$ events on all of the detector surface with $100$\% efficiency in this test-stand measurement, while incurring 0.2\% bulk event sacrifice. Both discriminators are expected to perform even more efficiently in the low-noise environment of the \MJ\ \DEM. Using the combined discriminator technique, we achieve a full order-of-magnitude improvement in the bulk event acceptance rate over the approach using only the rising-edge discriminator \cite{GERDA2017}. 

Applying the DCR technique in the \MJ\ \DEM, we find that the background rate in the \nonubb\ ROI is reduced by an order of magnitude. Given the observed position dependence of the reconstructed $\alpha$ energy, the spectrum of rejected $\alpha$ events should allow us to identify the source of $^{222}$Rn backgrounds with appropriate simulations in an upcoming full background model of the \MJ\ \DEM. The \MJ\ Collaboration is also developing an improved version of the DCR discriminator that corrects for bulk-trapping effects, which is beyond the scope of this paper. Efforts have also begun to study whether deep learning techniques can be used to identify the pulse shapes associated with $\alpha$ events. These new methods, along with the DCR technique, will be used to identify passivated surface events in other detector geometries, like the inverted-coaxial p-type point contact detectors planned for use in the upcoming LEGEND experiment.

\begin{acknowledgements}
We thank Tobias Bode for his many contributions to the measurement setup and our LEGEND colleagues for helpful discussions on charge collection behavior in \ppc\ detectors and $\alpha$ event analysis tools.

This material is based upon work supported by the U.S.~Department of Energy, Office of Science, Office of Nuclear Physics under contract/award numbers DE\hyp{}AC02\hyp{}05CH11231, DE\hyp{}AC05\hyp{}00OR22725, DE\hyp{}AC05\hyp{}76RL0130,
DE\hyp{}FG02\hyp{}97ER41020, DE\hyp{}FG02\hyp{}97ER41033, DE\hyp{}FG02\hyp{}97ER41041, DE-\-SC0012612, DE\hyp{}SC0014445, DE\hyp{}SC001\hyp{}8060, and LANLEM77. We acknowledge support from the Particle Astrophysics Program and Nuclear Physics Program of the National Science Foundation through grant numbers MRI-\-0923142, PHY\hyp{}1003399, PHY\hyp{}1102292, PHY\hyp{}1206314, PHY\hyp{}1614611, PHY\hyp{}1812409, and PHY\hyp{}1812356.  We gratefully acknowledge the support of the Laboratory
Directed Research \& Development (LDRD) program at Lawrence Berkeley National Laboratory for this work. We gratefully acknowledge the support of the U.S.~Department of Energy through the Los Alamos National Laboratory LDRD Program and through the Pacific Northwest National Laboratory LDRD Program for this work.  We acknowledge support from the Russian Foundation for Basic Research, grant No.~15\hyp{}02\hyp{}02919. We acknowledge the support of the Natural Sciences and Engineering Research Council of Canada, funding reference number SAPIN\hyp{}2017\hyp{}00023, and from the Canada Foundation for Innovation John R.~Evans Leaders Fund. This work was funded in part by the Deutsche Forschungsgemeinschaft (DFG, German Research Foundation) under Germany's Excellence Strategy \hyp{} EXC\hyp{}2094 \hyp{} 390783311. This research used resources provided by the Oak Ridge Leadership Computing Facility at Oak Ridge National Laboratory and by the National Energy Research Scientific Computing Center, a U.S.~Department of Energy Office of Science User Facility. We thank our hosts and colleagues at the Sanford Underground Research
Facility for their support. 

M.~Willers gratefully acknowledges support by the Alexander von Humboldt Foundation. J.~Gruszko gratefully acknowledges support of this work by the Pappalardo Fellowship, and the National Science Foundation Graduate Research Fellowship under Grant No.~1256082.

\noindent
{\bf Data Availability Statement} The results presented here are summarized in the form of histograms or appropriate graphs. The raw and processed data for these reported results (e.g. the waveforms or bin contents of the reported histograms) can be can be made available by the authors upon reasonable request.

\end{acknowledgements}

\appendix

\section{Scanner Engineering and Operation}\label{app:scanner}
The scanner (seen in Fig.~\ref{fig:TUBE}) consists of three main parts: the cryostat, detector holder, and collimator assembly. Details of the design of each aspect of the system are addressed here. 

The cryostat features a rail system that is mounted at the top of the vessel, with a rotational feedthrough on the side wall that allows the collimator radial position to be changed while the system is under vacuum. The collimator assembly is mounted to the carriage of this rail system, which has a pitch corresponding to 1.5\,mm of travel for every turn of the spindle. A 6-mm ``blind spot" on the detector surface is occluded by the contact pin and narrow plastic (polytetrafluoroethylene, or PTFE) holder that also provides routing for the signal cable running from the contact pin to the front-end electronics. 

The source position was set by hand and recorded throughout data-taking. Using this method, the uncertainty in step size between scanned positions is less than 0.3\,mm. The absolute position of the source is determined by taking data with the source at half-turn (0.75\,mm) intervals near the edges of the detector. The $\alpha$ event rate drops to 0 when the source beam is no longer incident on the detector surface, and the results of these measurements are combined with the known detector geometry to determine the source position on the detector surface. Given the 1.8\,mm beam spot size, a finer-grained scan would not improve the position determination. The resulting uncertainty in the source position is 0.75\,mm. 

\ppc\ detectors are highly infrared-shine sensitive, particularly with respect to their passivated surfaces; therefore the cryostat features multiple layers of IR shielding. The detector is housed inside a copper infrared (IR) shield (called the ``IR cup") with a 3\,mm-wide slit running along its diameter. This slit defines the axis that is scanned along, as the source beam shines through it onto the detector surface. Further shielding is provided by a cooled copper ``IR umbrella" shield, mounted on the tip of the collimator and moving along with the source. 

The IR cup shield is held at the ground potential, and is located 10\,mm above the detector's passivated surface. The combination of the IR cup and umbrella shields restricts the vacuum conductance into the area immediately surrounding the detector; though the vacuum level of the cryostat was measured to be $2.4\times10^{-6}$\,mbar, it is possible that the vacuum at the detector itself may have been worse by an order of magnitude. 

The source used is a 40\,kBq $^{241}$Am open source, with an expected full width at half maximum for the 5.486\,MeV $\alpha$ peak of less than 20\,keV. It is held in a 53\,mm-long collimator and suspended from the carriage of the rail system. The collimator bore is made of PTFE plastic, with a 10\,mm-thick copper end-cap, seen in Fig.~\ref{fig:TUBE}. The collimator is held at an incidence angle of $65\degree$, measured relative to the plane of the detector surface. The spot size of the source on the detector surface, which is calculated based on the known scanner and collimator geometry, is $1.8\pm0.1$\,mm in diameter.  

Given the source strength and collimator geometry, a source rate of 18\,mHz (65 events/hour) is expected at the detector surface. 84.8\% of these events, corresponding to a rate of 15\,mHz (54 events/hour) should include a 5.486\,MeV $\alpha$ emission. Another 13.1\% of events include a 5.443\,MeV $\alpha$ emission \cite{nudat_Am}. If the energy resolution of the $\alpha$ events is sufficiently reduced by interactions in the passivated surface, these peaks will be indistinguishable, and 98\% of the total activity will lie in the observed peak, for an expected signal rate of 17.6\,mHz.

Since the scanner is operated near sea-level (480\,m elevation), the expected cosmic muon rate is approximately 620\,mHz \cite{PDG}. Given the low $\alpha$ source event rate, an active veto system is used to reduce the cosmic muon background rate. It consists of a $82 \times 60 \times 5.5$\,cm-thick plastic scintillator panel coupled to a 2"-diameter photomultiplier tube, placed on top of the cryostat. This reduces the muon background rate to 175\,mHz in the 2.7 to 6\,MeV energy region where muons dominate the ambient background for these measurements.

\section{Energy Calibration and Stability}\label{app:calibration} 
Data from $\gamma$ calibration runs and $\alpha$ scan runs are processed with the same basic analysis pipeline. Waveforms are base\-line-sub\-tracted using the average of the first 500 samples (5\,$\mu$s) to calculate the baseline offset. They are pole-zero corrected for the dominant exponential decay of the pulses, driven by the resistive feedback preamplifier. The needed pole-zero correction is calculated by fitting an exponential decay to the tail of 1000 pulses in the 2615\,keV peak of a $^{228}$Th calibration run. The energy of each pulse was calculated from the maximum energy of a trapezoidal filter with an integration time of 8\,$\mu$s and a collection time of 3\,$\mu$s. The long collection time is chosen to avoid potential low-energy tailing in energy due to ballistic deficit, and also leads to a stable energy scale in spite of the drift time instability discussed in~\ref{app:stability}.

To calibrate the energy spectrum, fourteen peaks with energies between 295\,keV and 2615\,keV are fit simultaneously with a Gaussian peak, low energy tail, negligible high-energy tail, and step function centered at the Gaussian centroid. This is the same peak shape function used to fit the energy spectrum in the \MJ\ \DEM\ analysis; see Appendix A of Ref.~\cite{Guinn_Thesis} for details concerning the peak shape and the fitting procedure. 

The Gaussian peak centroids (in keV) are constrained to a quadratic function (in ADC), which is used to calibrate the energy scale. The quadratic component in the fit function corrects for small (less than 1\,keV) nonlinearities in the digitizer response \cite{Dolde2017}. The remaining non-linearity was measured by re-fitting the peaks with the peak centroids floated independently from one another; all peaks were fit to within 0.5\,keV of their original positions. The average FWHM at the 2615\,keV peak for all data sets is $3.2\pm0.6$\,keV. This resolution is similar to that obtained during the initial detector characterization (see Table~\ref{tab:PONaMA_specs}), and no attempts were made to further optimize the energy resolution.

The energy stability, which was evaluated on a run-by-run basis using ambient background peaks, was found to be good, with an average shift in the 2615\,keV peak of $0.68\pm0.60$\,ADC, corresponding to $0.92\pm0.81$\,keV. A single larger gain jump occurred halfway through data taking, likely due to changing environmental conditions (particularly room temperature) in the lab. All analysis parameters were re-calibrated following this gain change to avoid increased uncertainty due to the instability. 

\section{\texorpdfstring{$\alpha$}{Alpha} Spectral Peak Fits}\label{app:energy}
Events from the $\alpha$ scanning runs are processed with the same analysis pipeline used for $\gamma$ calibration events. The analysis cuts applied for spectral analysis and the $\alpha$ peak shape fitting functions used vary depending on the incident radius of the $\alpha$ events. 

\begin{figure}%
 \centering
 \subfloat[][Spectra taken without the $\alpha$ source incident on the passivated surface \textit{(in blue)}, and with the source incident at r=18.0\,mm \textit{(in red)}. The $\alpha$ peak is fit with a Gaussian function \textit{(see inset)}.]{
 \includegraphics[trim={0.6in 2.7in 1.2in 3in},clip,width=\linewidth]{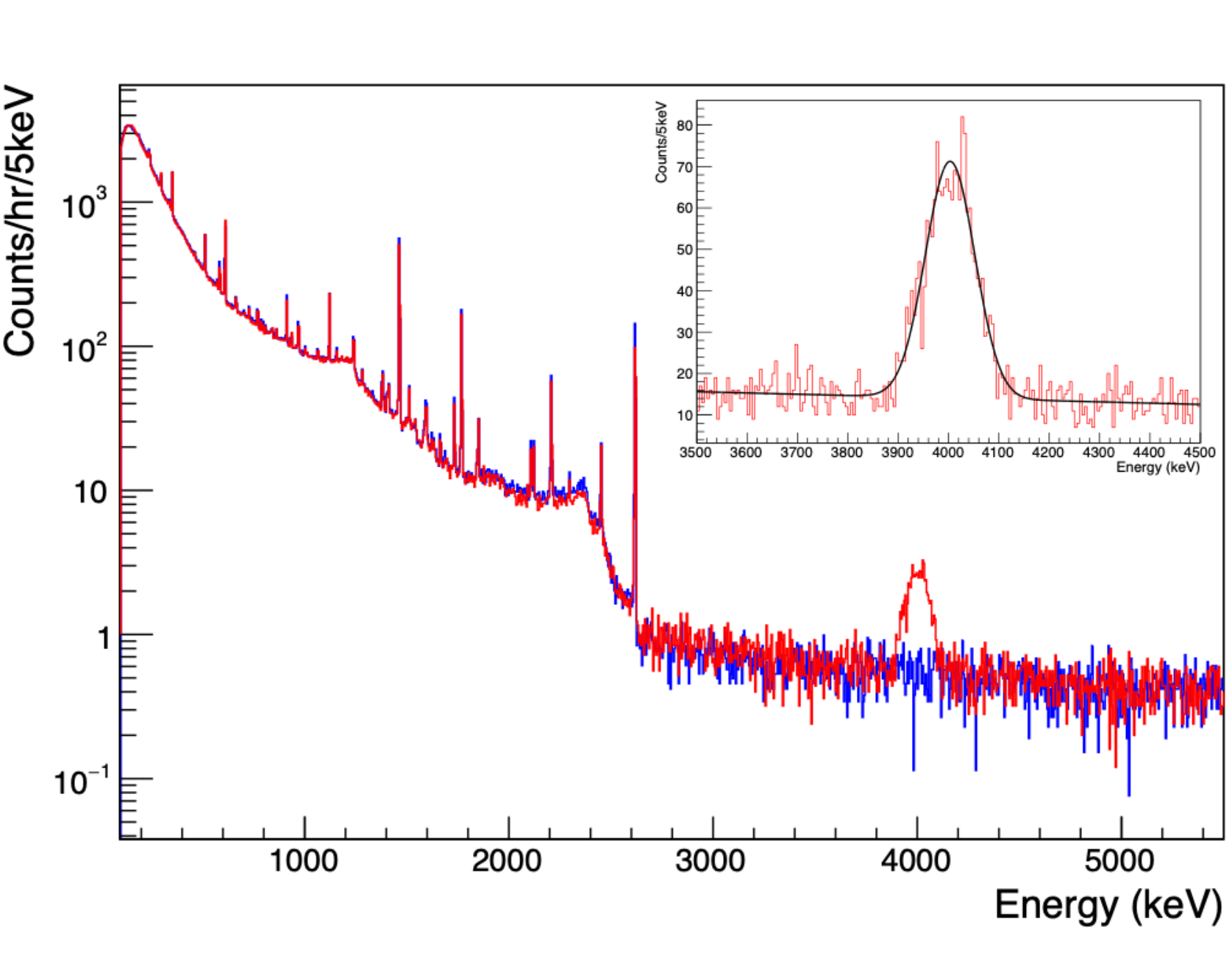}
 \label{fig:Espec}
 }
 \newline
 \subfloat[][Spectra taken with the $\alpha$ source incident on the passivated surface at $r=-4.5$\,mm \textit{(in red, left inset)}, and with the source incident on the p$^+$ contact ($r=-0.75$\,mm) \textit{(in blue, right inset)}. The $\alpha$ peaks are fit with the combination of a Gaussian and low-energy tail \textit{(see insets)}.]{
 \includegraphics[trim={1in 0.2in 0.5in 0.8in},clip,angle=270, origin=c, width=\linewidth]{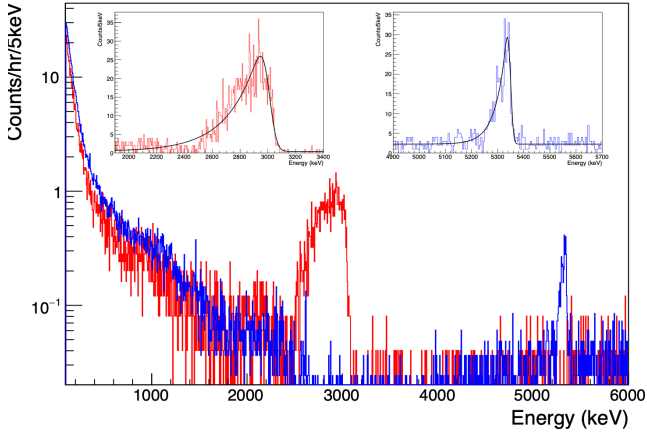}
 \label{fig:Espec_AEcut}}
 \caption{TUBE energy spectra and fits to $\alpha$ peaks \textit{(in black)}, with \textit{(bottom)} and without \textit{(top)} an additional cut to select near-point-contact events.} 
 \label{fig:Epeaks}
\end{figure}

For positions with radii larger than 6\,mm in magnitude, the mean $\alpha$ energy is larger than 2615\,keV, limiting the $\gamma$-interaction background contribution in the peak region. See Fig.~\ref{fig:Espec} for an example. Therefore, despite the low $\alpha$ interaction rate, the peak can be clearly identified and fit with a Gaussian peak and linear background spectral component that accounts for surviving muon and pile-up events, as shown in the inset of Fig.~\ref{fig:Espec}. No cut in A/E is used to select near-point-contact events in these data sets; it is not necessary to clearly distinguish the $\alpha$ event peaks in the energy spectra, and would in fact exclude the $\alpha$ event population at the largest scanning radii.

At radii smaller than 6\,mm, the $\alpha$ peak falls in an energy region of high ambient background rates. Due to the low $\alpha$ rate, the peak position cannot be identified without applying a pulse-shape cut to select the source events. A cut of A/E $>1.5$ selects near-point contact events (i.e.~those from the $\alpha$ source) while rejecting 99.8\% of background events. The $\alpha$ event peak is highly non-Gaussian, as in Fig.~\ref{fig:Espec_AEcut}, because of the wide range of weighting potential values sampled by the source beam width at small radii. In these peaks, there is significant low-energy tailing, and an exponentially-modified Gaussian tail is included in the fit (see left inset of Fig.~\ref{fig:Espec_AEcut}). 

In this case, the full function used to fit the peak is:
\begin{equation}
\begin{split}
    n(E) &= \frac{A(1-f_T)}{\sqrt{2\pi}\sigma}\text{Exp}(-\frac{(E-\mu)^2}{2\sigma}) \\
    &+ \frac{A f_T}{2\tau}\text{Exp}(\frac{\sigma^2}{2\tau^2} - \frac{E-\mu}{\tau})\text{Erfc}(\frac{\sigma}{\sqrt{2}\tau}-\frac{E-\mu}{\sqrt{2}\sigma}) \\
    &+ mE + b
    \end{split}
\end{equation}
where $A$ is the amplitude (i.e.~the total number of counts in the Gaussian and tail functions), $\mu$ and $\sigma$ are the mean and standard deviation, respectively, of the Gaussian function, $f_T$ is the fraction of the amplitude taken up by the low-energy tail, and $\tau$ is the decay constant of the tail exponential. $m$ and $b$ are the linear proportionality constant and flat portion of the background, respectively. $f_T$ is fixed at 0 for data sets with $r\geq6$\,mm, which show no low-energy tailing. More details concerning the peak shape function and fitting procedure can be found in Ref.~\cite{Guinn_Thesis}.

At small radii, the source beam width spans a large range of weighting potential values. In other words, the source spot size is large relative to the scale of the potential gradient in the detector at these positions, and therefore samples a larger range of electric field conditions. This leads to smearing of the energy peak, which appears as a significant low-energy tail. In these cases, the Gaussian peak width does not give an accurate energy range for the $\alpha$ events observed in these data sets. For scans with $|r|\leq4.5$\,mm, the estimated energy range of the observed $\alpha$ events given in Fig.~\ref{fig:EvR} is determined from the upper and lower energy bounds of the high-A/E event population that appears only in these $\alpha$ source runs. Since the events occur very near to the point contact, they have reliably high values of A/E. 

The event rate at these positions that are fully or partially-incident on the point contact is reduced due to the occlusion of the p$^+$ region by the contact pin. Again, an A/E cut selecting near-point-contact events (A/E$>1.5$) is applied to reduce the muon background rate, originally $0.1$\,events/hr/keV, by 2 orders of magnitude. The peak shape is well-approximated by the sum of a Gaussian and an exponentially-modified Gaussian. See the right inset of Fig.~\ref{fig:Espec_AEcut}. In this case, the low-energy tail of the peak is due to $\alpha$ energy loss in the 0.3\,$\mu$m-thick dead layer found at the p$^+$ contact surface.

\section{DCR Distribution Fits}\label{app:dcr}
The DCR of each event is determined as described in Sec.~\ref{ssec:PSD_DCR}. In all data sets with the source incident on the passivated surface, the distribution is fit with a Gaussian, and the underlying background events are fit with a step function centered at the mean of the Gaussian that accounts for the high-DCR tail of the bulk event distribution. The analysis cuts applied in fitting the resulting DCR distributions vary depending on the incident radius of the $\alpha$ events. 

\begin{figure}%
 \centering
  \subfloat[][The DCR distribution of single-site non-muon events with energies between 1 and 6\,MeV without the $\alpha$ source incident on the passivated surface \textit{(in blue)}, and with the source incident at r=18.0\,mm \textit{(in red)}.]{
 \includegraphics[trim={1in 0in 0in 0.5in},clip,angle=270, origin=c,width=\linewidth]{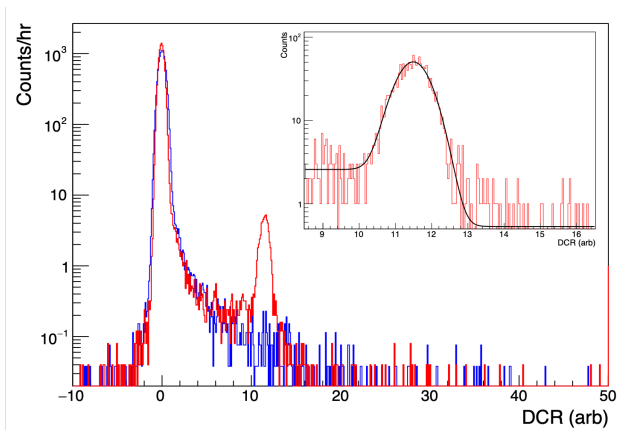}
 \label{fig:DCR}}
\hfill
 \subfloat[][The DCR distribution of single-site non-muon events with energies between 0.5 and 6\,MeV and A/E$>1.5$ without the $\alpha$ source incident on the passivated surface \textit{(in blue)}, and with the source incident at $r = 1.5$\,mm \textit{(in red)}. This source position also contributes events on the $p^+$ contact, which have higher energy and a DCR peak at 0. The $\alpha$ peak in DCR is associated with the energy-degraded events occurring on the passivated surface.]{
 \includegraphics[trim={1in 0in 0in 0.5in},clip,angle=270, origin=c,width=\linewidth]{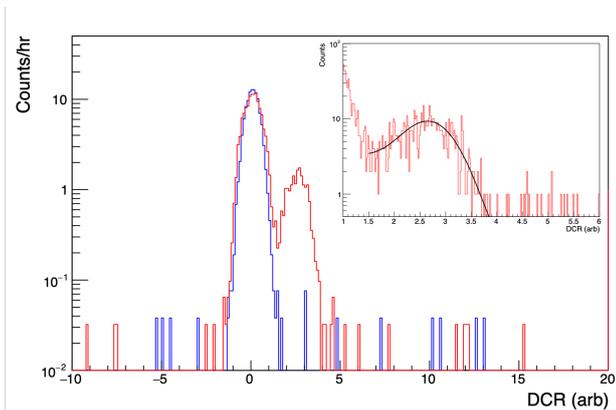}
 \label{fig:DCR_AEcut}}
 \caption{DCR parameter distributions and $\alpha$ peak fits. The peaks are fit with the sum of a Gaussian curve and a step function accounting for the tail of the bulk event distribution \textit{(in black, see insets)}.} 
 \label{fig:DCRpeaks}
\end{figure}

The DCR spectra used for scans with $|r| > 3$\,mm include all non-muon single-site events with energies between 1 and 6\,MeV. For $|r| > 3$\,mm, the energies of all observed $\alpha$ events fall in this range. Furthermore, the DCR value in the peak is sufficiently above the DCR distribution for normal events that the peaks can be clearly distinguished, and the high-DCR peak can be fit without any additional cuts (see Fig.~\ref{fig:DCR}).

For data sets with $|r|<3$\,mm, the relevant energy range extends below 1\,MeV when the low-energy tail of the $\alpha$ peak is taken into account. In these data sets the DCR values approach those of $\gamma$ background events, making the $\alpha$ event peak difficult to distinguish. As in the fits to the energy spectra, a pulse shape cut selecting near-point contact events (A/E $> 1.5$) is applied to reduce the background rate and allow a fit to the $\alpha$ events (see Fig.~\ref{fig:DCR_AEcut}). A broad energy range of 0.1 to 6\,MeV is included in the fit.

Events incident on the point contact itself do not have a distinct peak in DCR when the broad energy range of 1 to 6\,MeV is used. Instead, the peak must be fit using an energy window in which the $\alpha$ events dominate the spectrum; a 5$\sigma$ window around the peak energy is used. Due to the low event and background rate after these cuts are applied, the peak is fit using only a Gaussian distribution, with no underlying step function. This approach is used only for the data sets with $r=-0.75$\,mm, the source position at which the beam is entirely incident on the point contact.

\section{Charge Collection Instability}\label{app:stability}

\begin{figure}
  \centering
 \includegraphics[width=\linewidth]{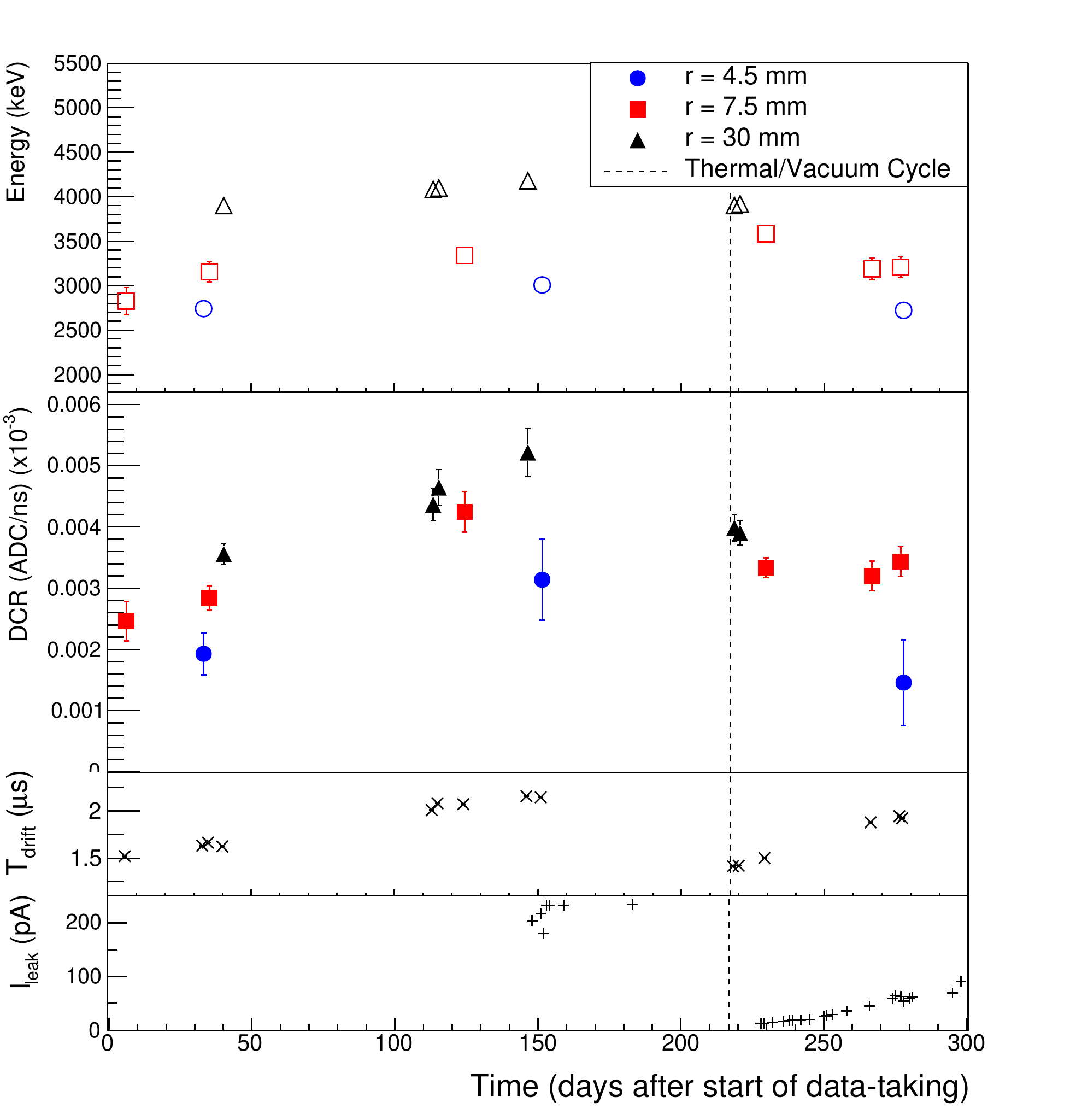}
 \caption{Stability studies of surface event charge collection conditions. The thermal and vacuum cycle date is indicated by the dashed line. \textit{Top:} Energies of the $\alpha$ peak with the source incident at several positions on the passivated surface, with positive and negative-radii scans combined. \textit{Second from top:} DCR values of the $\alpha$ peak in the same data sets. The error bars in the energy and DCR plots indicate the $3\sigma$ width of the peak. \textit{Second from bottom:} The long drift time peak centroid (see fit results in Fig.~\ref{fig:driftTime}) in each of the data sets displayed in the upper panels. \textit{Bottom:} The steady-state leakage current in the detector, measured at the test point of the 2002c Mirion preamplifier.}
 \label{fig:DCR_stability}
\end{figure}

\begin{figure}
  \centering
 \includegraphics[width=\linewidth]{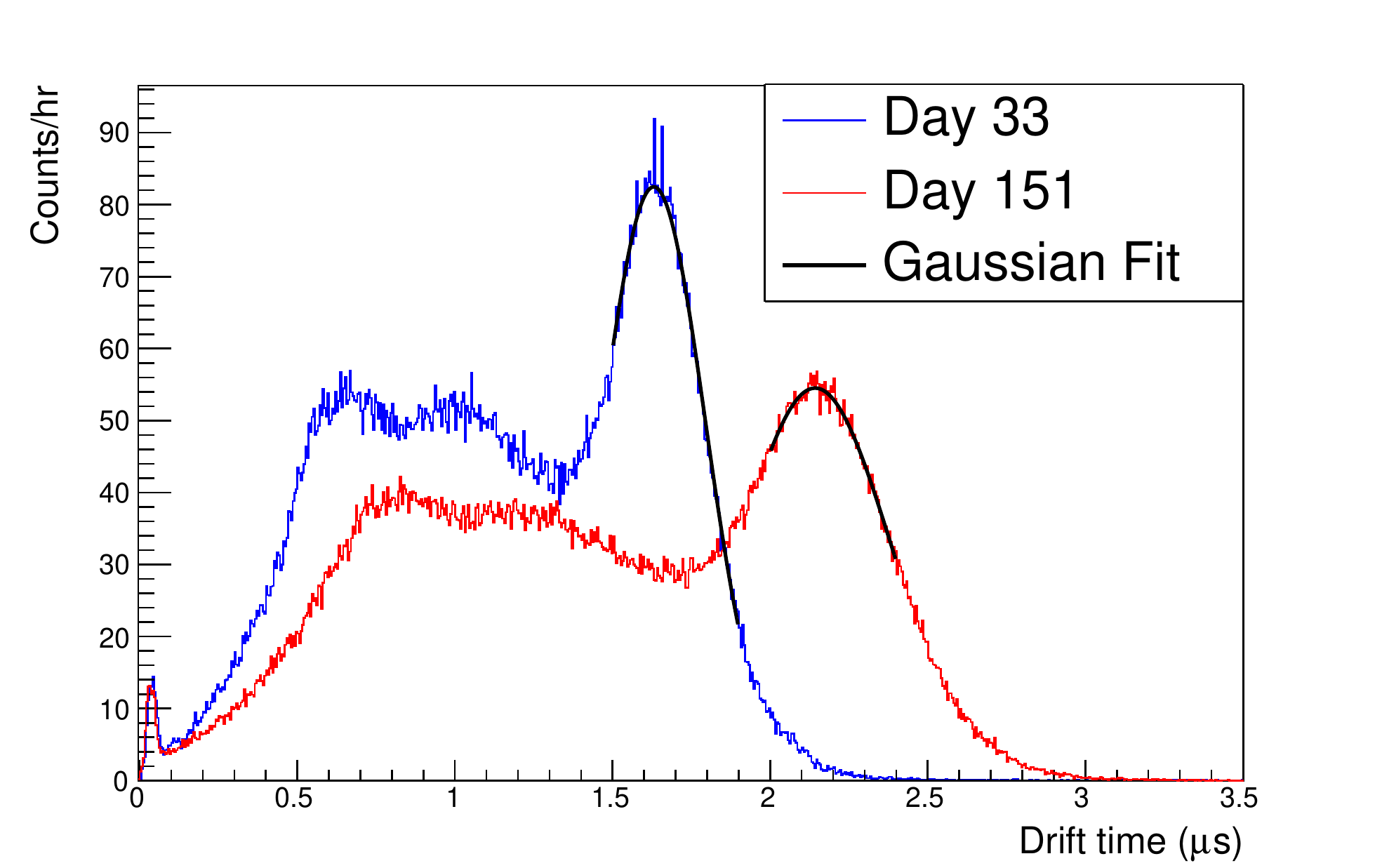}
 \caption{The drift time distribution for all events with energies between 1 and 6\,MeV during two data sets with the source incident at $r=-4.5$\,mm, taken after 33 days of data-taking \textit{(in blue)} and 151 days of data-taking \textit{(in red)}. The peak of (bulk) long drift-time events is fit with a Gaussian \textit{(in black)} to study the drift time stability over time. The small peak seen at drift times less than 100\,ns is due to the $\alpha$ events.}
 \label{fig:driftTime}
\end{figure}

Throughout the 9 months of data-taking with PONaMa-1 in the TUBE scanner, a variety of indications of instability in the charge collection conditions were observed. Though the drift time distribution of bulk $\gamma$ events was affected (see Fig.~\ref{fig:driftTime}), the detector gain and calibration peak pulse shapes are observed to be stable. This indicates that the bulk charge collection efficiency was not affected by the instability. The effect on surface $\alpha$ event energy and DCR, however, was significant.  

One candidate driver of the surface and near-p$^+$-contact charge build-up is trapped charge resulting from $\alpha$ interactions near the detector surface. Passivated surface charge build-up over time has been observed in other \ppc\ detectors \cite{Abt2017}, and our measurements of the observed energy indicate that significant charge is being lost on or near the surface. If this missing charge is trapped for timescales of several minutes or more, charge would accumulate on the surface even at the low (18 mHz) $\alpha$ irradiation rate used in this measurement.

Two other possible sources of instability are associated with poor vacuum conduction in the cryostat: charged ions may have adsorbed onto the surface of the detector, causing charge build-up, or the temperature of the detector may have risen over time due to degrading vacuum conditions. All three of these potential sources of instability would lead to increasing bulk event drift times over time.

A variety of radial positions were re-scanned after the 4\,months of initial data taking were complete to study the stability of the DCR response. In all of these measurements, the $\alpha$ events exhibited higher DCR values in the later data set. A thermal and vacuum cycle of the detector reduced the DCR values at all scanning positions, but it is unclear if the detector was completely restored to its original state, as the DCR values at some locations are larger than at the time they were first measured. The energy of the $\alpha$ peaks followed the same pattern, though with a smaller shift in magnitude, as seen in the top two panels of Fig.~\ref{fig:DCR_stability}. 

The rise in both observed energy and DCR is self-con\-sis\-tent; if trapped charge is being released more quickly, as indicated by the higher values of DCR, a larger fraction of that trapped charge will be collected promptly enough to contribute to the measured energy (within the $3\,\mu$s collection time of the trapezoidal filter). The full magnitude of the energy shift, however, cannot be accounted for by the change in the constant trapped charge release rate indicated by the shift in the DCR values. A non-constant component of the release rate may be responsible for the energy shift, or the fraction of the total charge that is trapped could be changing along with its release rate.  

The instability of the DCR parameter matches what would be expected to occur if passivated surface of the detector were becoming charged over time, whether due to ion adsorption onto the detector or the $\alpha$ irradiation. There were two additional indications of this charge build-up. The steady-state leakage current of the detector, monitored by measuring the leakage current test-point of the Mirion 2002C preamplifier, was 10\,pA before deploying the detector in the TUBE cryostat. Though the leakage current was not recorded regularly during the beginning of the first deployment, it was monitored after signs of DCR instability were found. It increased dramatically over the course of data-taking, as seen in the bottom panel of Fig.~\ref{fig:DCR_stability}, rising to 234\,pA before thermal/vacuum cycling. After cycling the system, it was restored to 12\,pA, and subsequently began to rise again.

Indications of surface charge build-up can also be seen in the drift time distribution for all events. As seen in Fig.~\ref{fig:driftTime}, the distribution, which under normal operating conditions is constant over time, became broader over the course of data-taking. The collection time of the trapezoidal filter was set to $3\,\mu$s, long enough to avoid incurring ballistic deficit even under the longest drift-time conditions. Therefore the bulk event energy and calibration peak shapes, including the energy resolution of the detector, were stable. The broadening of the drift time distribution, which corresponds to rescaling the original distribution by a constant, could indicate that positive charge is building up near the p$^+$-contact, slowing the portion of the signal created by electron-hole drift, which dominates the drift time of bulk events. 

To quantify the change in the drift time distribution over time, we fit the peak of long drift time events with a Gaussian, as shown in Fig.~\ref{fig:driftTime}. When the centroids are plotted in several data sets, as in Fig.~\ref{fig:DCR_stability}, they show a rise over time. The vacuum/thermal cycle of the system restored the drift times to their original values. The rate of rise in drift times during the second deployment was larger than in the first deployment. This is consistent with our hypothesis that surface and near-p$^+$-contact charge build-up driven by $\alpha$ interactions near the detector surface is responsible for the instability, since the $\alpha$ event rate was higher in the second deployment. 

Preliminary waveform simulations show that a change in the surface charge density of approximately $2\times10^9 e/cm^2$ would produce the observed effects. This quantity of charge could easily accumulate on the detector surface over the course of days or weeks if 1 to 10\% of the $\alpha$-induced charge carriers that are not collected promptly remain on the surface as trapped charge.

An increase in the detector temperature, due to poor vacuum conduction in the region immediately surrounding the detector, is an alternative potential cause of instability. This would similarly increase the bulk event drift times and the leakage current of the detector, as is observed. The DCR would also increase, with more thermal energy available to free charges from trapping sites and higher charge drift velocity at elevated temperatures both leading to a faster charge collection rate. This explanation would not motivate the increased instability following the thermal and vacuum cycle, however.

In lower $\alpha$ rate environments (such as the \MJ\ \DEM, which has an $\alpha$ event rate at least 6 orders of magnitude lower than the TUBE rate), the instability of the drift time and DCR parameter due to charge build-up is minimal. As is indicated by A vs.~E stability over the course of the \MJ\ \DEM\ data sets (see Fig. 7 in Ref.~\cite{MJD_avse}), no drift time rise over time is observed in the \MJ\ \DEM. This indicates that charge collection conditions in the detectors are stable. 

The additional measurements and simulations discussed in Section~\ref{sec:conc} should provide additional information on the origin of the observed instability in this and other test-stand measurements.

\section{\texorpdfstring{$\alpha$}{Alpha} Survival Probability Determination and Associated Uncertainties}\label{app:efficiency}
Save for the exceptions discussed here, the data sets are treated identically when determining their $\alpha$ survival probability.

The $\alpha$ survival probability at a given bulk event efficiency level is sensitive to the noise in the system and to the quality of the pole-zero correction, since these factors set the width of the bulk-event DCR distribution. The uncertainty in the efficiency is sensitive to the $\alpha$ event rate and any changes in the background event rate over the course of data-taking. The background $\gamma$ rate at and below 2614.5\,keV was far more variable than the cosmic $\mu$ rate above this energy. 

In two cases, the data sets taken at source positions of $r= -7.5$ and -4.5\,mm, the signal window is shrunk by 52 and 31\,keV, respectively, to avoid the $^{208}$Th 2614.5\,keV peak region. In both cases, the signal region is narrowed by less than $1\sigma$ in energy, and we still expect the efficiency calculation to include over 90\% of $\alpha$ peak events. In these cases, including the $\gamma$ background peak in the signal region unnecessarily increases the statistical uncertainty of the efficiency determination.

In the data set with the source incident at $r=-2.25$\,mm, the calculated $\alpha$ event rate is found to be highly unphysical, which indicates that there was a significant excess of background events in the source-free run compared to the rate expected based on the relative live-times. The partial occlusion of the source beam leads us to expect a small number of $\alpha$ events at this position (about 150 in total), and their large energy range (see Fig.~\ref{fig:EvR}) and the high ambient backgrounds at these energies make relatively large fluctuations in the background rate likely. The $\alpha$ rejection efficiencies are not given for this position, and are not included in the average rejection values. 

\bibliographystyle{spphys} 
\bibliography{main.bib}

\end{document}